\documentclass[a4paper, 12pt]{article}
\RequirePackage[OT1]{fontenc}
\usepackage{amsfonts,amssymb,graphics,epsfig,verbatim,bm,latexsym,amsmath,url,amsbsy,arydshln,multirow,rotating,enumerate,graphicx,pgf,tikz,color,xcolor,lscape,amstext,dsfont,caption,subcaption,colortbl,scalerel,stackengine, csquotes,verbatim,setspace}
\usepackage[textheight=722pt,textwidth=470pt,centering,headheight=50pt,headsep=12pt,footskip=12pt]{geometry}
\usepackage{tablefootnote}
\usepackage[colorlinks,citecolor=purple!60!black,linkcolor=blue!60!black,urlcolor=purple!60!black]{hyperref}
\usepackage[framemethod=TikZ]{mdframed}
\usepackage{longtable}  
\usepackage{tabularx}   %
\usepackage{rotating}  

\usepackage{setspace}
\usepackage{caption}
\usepackage{subcaption}
\usepackage{adjustbox} 
\usepackage{graphics}
\usepackage[numbers, sort&compress]{natbib}

\usepackage{blindtext}
\usepackage[T1]{fontenc}
\usepackage[ngerman, english]{babel}
\usepackage[titletoc, page]{appendix}

\usepackage{float} 

\usepackage{tikz}
\usetikzlibrary{shapes.geometric}%
\usetikzlibrary{arrows,decorations.markings}

\usepackage{footmisc} 

\newcommand\independent{\protect\mathpalette{\protect\independenT}{\perp}}
\def\independenT#1#2{\mathrel{\rlap{$#1#2$}\mkern2mu{#1#2}}}

\makeatletter
\newcounter{step}
\renewcommand*{\thestep}{Question~\arabic{step}}

\newcommand*{\REF}[1]{%
  \begingroup
    \refstepcounter{step}%
    \IfRefUndefinedBabel{step:#1}{%
    }{%
      \edef\@currentHref{%
        \getrefbykeydefault{step:#1}{anchor}{}%
      }%
    }%
    \label{#1}%
    \hyperref[{step:#1}]{\thestep}%
  \endgroup
}

\newcommand*{\STEP}[1]{%
  \begingroup
    \phantomsection
    \label{step:#1}%
    (\ref*{#1})%
  \endgroup
}
\makeatother

\makeatletter
\newcounter{assumption}
\renewcommand*{\theassumption}{A\arabic{assumption}}

\newcommand*{\ass}[1]{%
  \begingroup
    \refstepcounter{assumption}%
    \IfRefUndefinedBabel{assumption:#1}{%
    }{%
      \edef\@currentHref{%
        \getrefbykeydefault{assumption:#1}{anchor}{}%
      }%
    }%
    \label{#1}%
    \hyperref[{assumption:#1}]{\theassumption}%
  \endgroup
}

\newcommand*{\ASSUMPTION}[1]{%
  \begingroup
    \phantomsection
    \label{assumption:#1}%
    \ref*{#1}%
  \endgroup
}
\makeatother

\usepackage{makecell}[l]

\begin{document}
\begin{center}
\textbf{{\LARGE{Targeting Mediating Mechanisms of Social Disparities with an Interventional Effects Framework, Applied to the Gender Pay Gap in Western Germany}}}
\end{center}\vspace*{0.5cm}

{\Large
\begin{center}
Christiane Didden\footnote{Department of Sociology, LMU Munich, Munich, \href{mailto:Christiane.Didden@soziologie.uni-muenchen.de}{Christiane.Didden@soziologie.uni-muenchen.de}
}

\end{center}
}

\begin{abstract}

The Oaxaca-Blinder decomposition is a widely used method to explain social disparities. However, assigning causal meaning to its estimated components requires strong assumptions that often lack explicit justification. This article emphasizes the importance of clearly defined estimands and their identification when targeting mediating mechanisms of social disparities. Three approaches are distinguished on the basis of their scientific questions and assumptions: a mediation approach and two interventional approaches. The Oaxaca-Blinder decomposition and Monte Carlo simulation-based g-computation are discussed for estimation in relation to these approaches. The latter method is used in an interventional effects analysis of the observed gender pay gap in Western Germany, using data from the 2017 German Socio-Economic Panel.
Ten mediators are considered, including indicators of human capital and job characteristics. Key findings suggest that the gender pay gap in log hourly wages could be reduced by up to 86\% if these mediators were equally distributed between women and men. Substantial reductions could be achieved by aligning full-time employment and work experience.

\end{abstract}

\section{Traditional and modern approaches to decomposing social disparities}
\label{Introduction}
Social disparities exist in various domains, including wealth, education, the labor market, health, and political participation \citep{neckerman2004social}. 
A prominent example is the gender pay gap, which typically indicates that women, on average, earn less than men. Gender differences in wage-relevant factors, such as education, career paths, work experience, and working hours, are typically used to explain this phenomenon \citep{Bobbitt_2007,  Golding_2014,  Blau_Kahn_2017, kunze2018gender, schmitt_stall_2022}. A common method to evaluate the contribution of these explanatory factors to wage disparities between two groups, also applied by official statistical offices \citep{Mischler2021Verdienstunterschiede}, is the Oaxaca-Blinder (OB)
decomposition method \citep{Oaxaca1973, Blinder1973}. Central to this method is the twofold decomposition for linear outcome models. A linear regression model with wage as the outcome is fitted for each of the two groups with the explanatory factors as regressors. Based on these models, the marginal wage disparity is then decomposed into two components: the ``explained'' part,  which relates to the differences between the two groups in the means of the explanatory variables, and the ``unexplained'' part,  which relates to the differences between the two groups in the associations between the explanatory variables and wage. 
 This relatively straightforward decomposition, implemented in common software packages \citep[][]{jann2008blinder, hlavac2018oaxaca}, is not restricted to wage gaps, but can be applied to various outcome disparities (see, e.g., \citep[][]{sen2014using}). 
 
\begin{figure}[H]

 \centering
\subfloat[]{\label{Structural assumptions to illustrate Approaches 1-3}
 \centering
 \scalebox{0.9}{
\begin{tikzpicture}[x=10in,y=6in]
\node (M1) at (0.15,-0.050) {$M_1 \cdots M_{k-1}$};
\node (A) at (-0.02,-0.300) {$A$};
\node (M7) at (0.15,-0.150) {$M_k$};
\node (M10) at (0.15,-0.250) {$M_{k+1} \cdots M_{K}$};
\node (Y) at (0.3,-0.300) {$Y$};
\node (C) at (-.05,-0.0) {$\boldsymbol{C}$};
\node (L) at (0.07,0.0) {$\boldsymbol{L}$};

\draw [->] (A) edge (M1);
\draw [->] (A) edge (M7);
\draw [->] (A) edge (M10);
\draw [->] (M1) edge (Y);
\draw [->] (M7) edge (Y);
\draw [->] (M10) edge (Y);
\draw [->] (C) edge (M1);
\draw [->] (C) edge (M7);
\draw [->] (C) edge (M10);
\draw [->] (C) edge (Y);
\draw [->] (C) edge (A);
\draw [->] (A) edge (Y);
\draw [->] (A) edge (L);
\draw [->] (L) edge (M1);
\draw [->] (L) edge (M7);
\draw [->] (L) edge (M10);
\draw [->] (L) to [out=20,in=90, looseness=0.8] (Y);

\draw [->] (C) edge (L);

\draw [loosely dotted, very thick]  (M1) edge (M7);
\draw [loosely dotted, very thick] (M7) edge (M10);
\draw [loosely dotted, very thick] (M1) to [out=-45,in=45, looseness=0.5] (M10);

\end{tikzpicture}
}
}\vspace*{0.8cm}
\hspace*{0.8cm}
   \subfloat[]{
\label{Graph2}

 \centering
 \scalebox{0.8}{
\begin{tikzpicture}[x=6in,y=6in]
\node (A) at (0.000,-0.300) {$A$};
\node (M) at (0.25,-0.150) {$M$};
\node (Y) at (0.5,-0.300) {$Y$};
\node (C) at (-.05,-0.07) {$C$};

\node (L) at (0.1,0.0) {$L$};

\draw [->] (A) edge (M);
\draw [->] (M) edge (Y);
\draw [->] (C) edge (M);

\draw [->] (C) edge (A);
\draw [->] (C) edge (Y);

\draw [->] (A) edge (Y);
\draw [->] (A) edge (L);
\draw [->] (L) edge (M);
\draw [->] (C) edge (L);
\draw [->] (L) edge (Y);

\end{tikzpicture}
}}
\subfloat[]{\label{Graph1}
 \centering
 \scalebox{0.8}{
\begin{tikzpicture}[x=6in,y=6in]

\node (A) at (0.000,-0.300) {$A$};
\node (M) at (0.25,-0.150) {$M$};
\node (Y) at (0.5,-0.300) {$Y$};
\node (C) at (-.05,-0.07) {$C$};
\node (L) at (0.1,0.0) {$L$};

\draw [->] (A) edge (M);
\draw [->] (M) edge (Y);
\draw [->] (C) edge (M);

\draw [->] (C) edge (Y);

\draw [->] (A) edge (Y);
\draw [->] (A) edge (L);
\draw [->] (L) edge (M);
\draw [->] (C) edge (L);
\draw [->] (L) edge (Y);
\end{tikzpicture}
}

}
       
        \caption{Structural assumptions: (a) with multiple mediators $M_1, \ldots, M_K$, (b) and (c) with a single mediator $M$. Dotted lines among mediators indicate unknown structural dependencies among them. 
    } 
        
\end{figure}
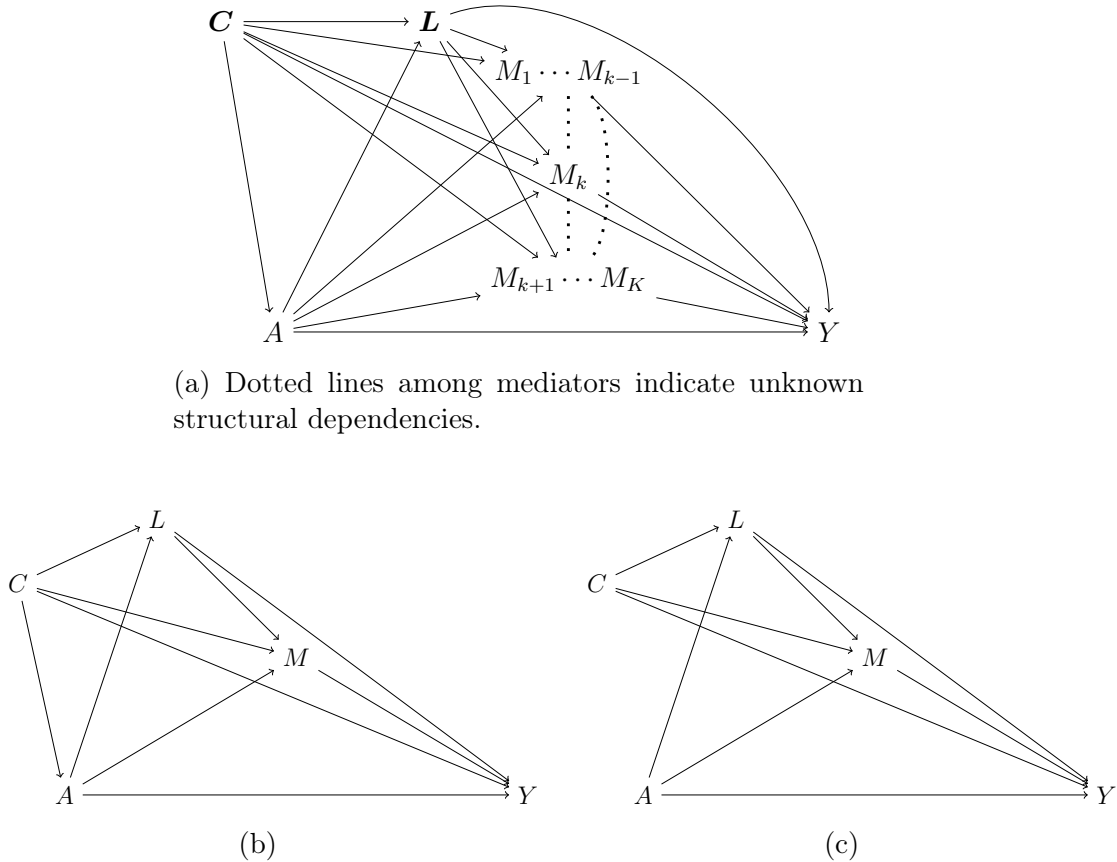

Decomposition analyses that aim to provide explanations for disparities beyond mere statistical descriptions suggest a focus on causal relationships. 
\citet{Petersen_VanderLaan_Roadmap2014} and \citet{Lundberg_etal_2021} emphasized the importance of guiding a causal analysis with a clearly defined target parameter. The steps involved in a causal analysis using observational data, as outlined by \citet{Petersen_VanderLaan_Roadmap2014}, can be categorized into three stages: the conceptual, the statistical, and the interpretation stage.
The conceptual stage involves several key steps. First, a causal model is established to summarize assumptions about the data-generating process. The observed data are linked to this model, followed by a formulation of a precise research question. This question is then translated into a formal target parameter, known as the causal estimand. Finally, the identifiability of the parameter is evaluated; if it is identified, it can be expressed as a function of the observed data. The subsequent statistical stage involves specifying the statistical estimation problem, selecting the estimation method, and performing the estimation. The final stage focuses on selecting an appropriate interpretation of the estimated parameter.
According to this roadmap, the conceptual stage underpins the analysis and guides the subsequent steps, not the formulation of a statistical model. However, researchers have noted that despite this crucial role, conceptual foundations including the definition of a causal estimand are often neglected in social science applications \citep{Lundberg_etal_2021}, particularly in decomposition analyses, including the OB decomposition \citep{Huber2015, Fortin}. This article addresses this concern by explicitly emphasizing conceptual aspects in research targeting mediating mechanisms of social disparities, incorporating recent methodological findings \citep{VanderWeeleVansteeland_Robins_RIA2014, Jackson, Daniel2017, Moreno_2021_emulate_target, NguyenSchmidOgburnStuart_2022,  Miles_2023}. Central to the paper is a nonparametric causal model that is adaptable to a wide range of social contexts. It includes a time-fixed exposure (or treatment), a single outcome, multiple interdependent mediators, and different types of confounders (see directed acyclic graph (DAG) in Figure \ref{Structural assumptions to illustrate Approaches 1-3}). The exposure variable pertains to a social characteristic that may be ascribed, such as gender or race, or may be acquired over the course of an individual's lifetime, such as educational attainment or union membership. It is assumed to affect the outcome both directly and indirectly through intermediate factors, i.e., mediators. Drawing on recent classifications of effect types \citep{nguyen_clarifying_2021} and methodological insights \citep{Miles_2023}, three approaches are distinguished. Although all three approaches target mediating mechanisms of social disparities, they differ in terms of their scientific questions, estimands, and the strength of the required identification assumptions.
The first approach (Approach 1) aims to conduct a causal mediation analysis to explain how the exposure affects the outcome through direct and indirect pathways (see \citep[][]{VanderWeeleVansteeland_Robins_RIA2014, Daniel2017, Miles_2023}, among others). The second approach (Approach 2) is an interventional approach that aims to evaluate the effects of hypothetical mediator manipulations on the exposure-induced outcome disparity \citep[][]{Moreno_2021_emulate_target}. The third approach (Approach 3) is an interventional approach that aims to evaluate the effects of manipulations of explanatory factors, such as potential mediators, on the actual observed outcome disparity \citep[][]{vanderweelerobinson2014causalrace, Jackson}.

Approaches 1–3 will be illustrated by nonparametrically defined causal estimands. These quantities do not depend on specific model assumptions or parameters, unlike the effects in traditional parametric methods in mediation \citep{judd1981process, baron1986moderator} (see \citep{Valerie_Interac, Shpitser2013} for limitations of these methods). Specifically, causal effects are defined on the additive scale as contrasts between two potentially counterfactual mean outcomes: one that would be observed in response to a specific intervention and another that would be observed without this intervention (or in response to a different intervention) \citep[][]{VanderWeele_2016_Comment_Causes}. The term ``intervention'' is employed in a hypothetical manner to describe a potential manipulation of a variable, without specifying the details of how this intervention could be implemented or by whom. With regard to the mediator interventions considered, this concept aligns with the ill-defined type of interventions discussed in \citet{Moreno_2021_emulate_target}.
As noted by \citet{nguyen_clarifying_2021}, any contrast of outcomes under two different intervention conditions that could be conceivable for hypothetical future studies falls under the class of interventional effects. This article focuses on a specific type of interventional effect, namely, that which is characterized by interventions on the distribution of hypothesized mediators. This type of interventional effect has received substantial attention in methods research on mediation analysis (see, e.g., \citep[][]{Didelez2006, VanderWeeleVansteeland_Robins_RIA2014, Daniel2017, Tylervanderweele_Tchetgen_2017mediation, Miles_2023}). Applications can be found in the field of epidemiology, in particular (see, e.g., \citep[][]{blaikie2023differential, rudolph2021helped, diddencontribution_2024}).

Causal implementations of the twofold OB decomposition and the associated identification assumptions have been discussed in previous research with different foci. \citet{Fortin} focused on causal inferences more generally, with a particular emphasis on the interpretation of the unexplained component. \citet{Huber2015} examined the method within the context of mediation analysis. \citet{Jackson} linked the twofold OB method to an interventional approach for observed disparities. This article contributes by examining the applicability of the twofold OB decomposition for linear outcome models as an estimator in the context of Approaches 1-3. For simplicity, causal models with a single mediator are used for this purpose. Unlike the causal diagrams in \citet{Huber2015, Jackson}, these models include a confounder of the mediator-outcome relationship that is affected by the exposure and may interact with the mediator in its effect on the outcome.

Finally, the paper applies Approach 3 to the gender pay gap, using data from the 2017 German Socio-Economic Panel (SOEP) \citep{SOEP, goebel2019German} on households located in the region of former West Germany, hereafter referred to as Western Germany.
Ten potential mediators are considered including job characteristics as well as measures of education and labor market experience, both key components of human capital \citep{Becker_1962, Mincer_1974_schooling}.
The rationale for using a g-computation approach for estimation is outlined, and the steps involved in the Monte Carlo simulation-based g-computation, hereafter referred to as MC g-computation, are described. 

The structure of this article is as follows: it begins with an introduction to counterfactual notation (Section \ref{Notation}) and a general review of interventional effects defined by stochastic mediator interventions (Section \ref{Section: Interventional effects general}). Approaches 1-3 are then outlined and illustrated with specific causal estimands (Sections \ref{Approach1}, \ref{Approach2}, \ref{Approach3}). Section \ref{Section Oaxaca} addresses the use of the twofold OB decomposition for linear outcome models in the context of Approaches 1-3. Section \ref{Application}  illustrates the application of Approach 3 to the gender pay gap in Western Germany. The adaptability of the approach is demonstrated by considering different types of mediator interventions.
The article concludes with a discussion in Section \ref{Section Discussion}.

\tikzset{
  big arrow/.style={
    decoration={markings,mark=at position 1 with {\arrow[scale=4,#1]{>}}},
    postaction={decorate},
    shorten >=0.4pt},
  big arrow/.default=blue}

 \section{Counterfactual Notation}
 \label{Notation}
Let $A$ denote the exposure and $Y$ the outcome. Let $M$ denote an intermediate variable, which is assumed to mediate the effect of $A$ on $Y$. 
Let $\boldsymbol{C}$ denote a set of baseline covariates, including exposure-mediator, exposure-outcome, and mediator-outcome confounders that are not affected by exposure. Additionally, let $\boldsymbol{L}$ denote a set of intermediate mediator-outcome confounders that are affected by the exposure; therefore, they may also act as mediators.
  
 Let $Y_{a}$ denote the counterfactual outcome $Y$ under an intervention that sets the exposure to the value $a$. The expected counterfactual outcome under $A=a$, denoted by $E[Y_{a}]$, generally differs from $E[Y|A=a]$, the expected outcome at the actual exposure level $a$. This difference arises because $E[Y_{a}]$ represents what the expected outcome would be in a hypothetical scenario where the exposure value is $a$ for all individuals in the population of interest, while $E[Y|A=a]$ represents the expected outcome among individuals who are actually exposed to $a$. Central to mediation are counterfactuals of the form $Y_{a m}$, defined by interventions on the exposure and on the mediator. According to the composition assumption, $Y_a$ can be expressed as $Y_{a M_a}$, which is the counterfactual outcome under $A=a$ and the mediator value naturally arising under $A=a$ \citep[][]{pearl_2009}. Additionally, using the notation in \citet{loh2021heterogeneous}, let $\widetilde{M}_{a|\boldsymbol{C}}$ be a random draw from the counterfactual distribution of $M$ under $A=a$, given $\boldsymbol{C}$, denoted by $P_{M_{a|\boldsymbol{C}}}(m|\boldsymbol{C})$. Furthermore, let $\widetilde{M}|A=a,\boldsymbol{C}$ denote a random draw from the distribution of $M$ 
 among those with exposure value $a$ and covariates $\boldsymbol{C}$, given by $P_{M|A=a,\boldsymbol{C}}(m|A=a, \boldsymbol{C})$.\footnote{{$P_{M|A,\boldsymbol{C}}(m|a, \boldsymbol{C})=P(M=m|A=a, \boldsymbol{C})$, and $P_{M_a|\boldsymbol{C}}(m|\boldsymbol{C})=P(M_a=m|\boldsymbol{C})$.}}
{More generally, let $\widetilde{M}$ denote a random draw from a user-specified distribution of $M$.

The focus of Sections \ref{Approach1} - \ref{Approach3} and Sections \ref{Section Oaxaca} - \ref{Application} is on a binary exposure that takes values in $\{0, 1\}$.  In this context, $E[Y_{1}]$ refers to the expected counterfactual outcome \textit{under exposure}, and $E[Y_{0}]$ to the expected counterfactual outcome \textit{under control}.  $E[Y|A={1}]$ refers to the expected outcome among the actual observed exposure group (the exposed), and $E[Y|A={0}]$ to the expected outcome among the actual observed control group (the unexposed).

 \section{Interventional effects for decomposing social disparities}
 \label{Section: Interventional effects general}
Mediation is commonly understood as a phenomenon that occurs when the exposure affects a mediator $M$ and the resulting change in $M$ {goes on to} affect the outcome \citep{Tyler_book, Miles_2023}. 
The central measure for the indirect effect through $M$ is the natural indirect effect (NIE) through $M$, usually defined in the mediation literature by NIE$_{M}=E[Y_{a M_a}-Y_{a M_{a^*}}]$\footnote{{\citet{robins1992identifiability}} differ between the ``total indirect effect'', given by $E[Y_{a M_a}-Y_{a M_{a^*}}]$, and the ``pure indirect effect'', given by $E[Y_{a^* M_a}-Y_{a^* M_{a^*}}]$.\label{footnotePure}}{, with $a$ referring to the exposure level and $a^*$ to the control level \citep{Pearl2001}}. The effect of the exposure on the outcome that is not mediated by $M$ is known as the natural direct effect (NDE), defined as NDE$_{M}=E[Y_{a M_{a^*}}-Y_{a^*M_{a^*}}]$. Both the NIE$_{M}$ and NDE${_M}$ are measures of the underlying mechanisms through which the exposure affects the outcome. Notably, they add up to the total effect of the exposure  (TE), also known as average treatment effect, given by $\text{TE}=E[Y_{a M_a}-Y_{a{^*}M_{a^*}}]$ \citep{Pearl2001}. In this article, the TE is also referred to as the \textit{exposure-induced disparity}. 
Natural effects are defined starting at the individual level, with average natural (in)direct effects being population means of the individual (in)direct effects \citep{moreno2018understanding, nguyen_clarifying_2021}. However, they involve cross-world counterfactual outcomes in the form of $Y_{a M_{a^*}}$ that are empirically inaccessible because it is not possible to bring an individual's exposure to level $a$ and the mediator for that individual to its natural level under the counterfactual level $a^*$. Therefore, the identification of natural (in)direct effects requires strong assumptions, including the cross-world independence assumption $Y_{a m}\independent M{_{a^*}}| \boldsymbol{C}$ ($a\neq a^*$). This assumption does often not hold, as it is known to be violated in the presence of confounders of the mediator-outcome relationship affected by the exposure, referred to as \textit{exposure-induced confounders} \citep{Avin2005}. Even if a well-controlled randomized experiment were feasible, it cannot eliminate exposure-induced confounding of the mediator-outcome relationship \citep{Miles_2023}. In response to this problem, interventional (in)direct  effects, also described as ``randomized interventional analog[ues] of the notions of natural direct and indirect effects'' \citep{VanderWeeleVansteeland_Robins_RIA2014}, have been examined as alternatives to natural effects (see \citep[][]{VanderWeeleVansteeland_Robins_RIA2014, Daniel2017, Tylervanderweele_Tchetgen_2017mediation, Zheng_Laan_Survival_2017, Miles_2023}, among others). 
 Interventional (in)direct effects are defined by weighted expectations of counterfactual outcomes, weighted according to counterfactual mediator distributions representing specific stochastic mediator interventions \citep{moreno2018understanding}. In particular, the interventional analogue to NIE$_{M}$, as proposed by \citet{VanderWeeleVansteeland_Robins_RIA2014}, is given by 
 $\text{IIE}_{M|\boldsymbol{C}}=E[Y_{a \widetilde{M}_{a|\boldsymbol{C}}}-Y_{a \widetilde{M}_{a^*|\boldsymbol{C}}}]$ (see \citep{Tylervanderweele_Tchetgen_2017mediation} for an alternative interventional analogue\footnote{This alternative interventional analogue to NIE$_{M}$ is $E[Y_{a}-Y_{a \widetilde{M}_{a^*|\boldsymbol{C}}}]$, which compares the expected counterfactual outcome under $A=a$ with the expected outcome when the mediator is randomly drawn from its counterfactual distribution under $A=a^*$, given $\boldsymbol{C}$. The corresponding analogue to NDE$_{M}$, given by $E[Y_{a \widetilde{M}_{a^*|\boldsymbol{C}}}-Y_{a^*}]$, however, captures both the effect of changing the exposure from $a^*$ to $a$ and the effect of having the mediator set to its natural level under $a^*$ compared to a random draw from the mediator under $a^*$ \citep[][]{Tylervanderweele_Tchetgen_2017mediation}.}). This effect is defined by the contrast between (1) the counterfactual outcome expected under $A=a$ and the mediator being randomly drawn from its counterfactual distribution under $A=a$, given $\boldsymbol{C}$, and (2) the counterfactual outcome expected under $A=a$ and the mediator being randomly drawn from its counterfactual distribution under $A=a^*$, given $\boldsymbol{C}$. It follows that the contrast between these two expected counterfactual outcomes is due to a distributional shift in the mediator - from the distribution under the exposure level $a$ to that under the level $a^*$, given $\boldsymbol{C}$. These counterfactual \textit{mediator intervention distributions} are marginal with respect to exposure-induced confounders $\boldsymbol{L}$, ignoring the dependence between
$M$ and $\boldsymbol{L}$.  Notably, interventional (in)direct effects do not rely on cross-world independence assumptions, which is crucial in contexts with multiple interdependent mediators, where these assumptions are typically violated \citep{Tylervanderweele_Tchetgen_2017mediation, DíazWilliamsRudolph_Longitduinal_2023, Zheng_Laan_Survival_2017, Lin_Survival_2017}.

\citet{nguyen_clarifying_2021} note that discussing interventional (in)direct effects at the individual level is not meaningful because they depend on mediator values randomly drawn from counterfactual distributions in subpopulations defined by the covariate strata in $\boldsymbol{C}$. \citet{moreno2018understanding} argue that such ``population-level interventions'' could be conceptualized for hypothetical future trials. Therefore, interventional (in)direct effects may serve as ``more natural initial target estimands for mediation'' than natural (in)direct effects, which rely on infeasible ``individual-level interventions''. \citet[][]{Miles_2023}, however, emphasized the importance of the individual level in mediation analysis, arguing that any true effect measure for the mediated effect through $M$ ought to take its null value when no individual-level indirect effect through $M$ exists in the population of interest.
Certain indirect effect measure criteria, namely, the sharp(er) null criterion and the monotonicity criterion, ought to be satisfied when measuring mediation and the direction of the mediated effect.
To ensure that interventional indirect effects meet these criteria and can serve as substitutes for natural indirect effects - which satisfy these criteria - restrictive conditions must hold beyond those required for identification. Otherwise, interventional indirect effects may lack true mediational meaning, potentially showing non-zero values even in the absence of individual-level indirect effects. Furthermore, they may misidentify the direction of the mediated effect, even if it were uniform (or null) across all individuals \citep{Miles_2023}.
However, in the absence of mediational meaning, interventional effects - defined by interventions on hypothesized mediators - can still offer valuable insights into the extent to which social disparities could change if mediating mechanisms were altered \citep{nguyen_clarifying_2021, Moreno_2021_emulate_target, Jackson}.

Drawing from recent literature on mediation and interventional effects \citep{Miles_2023, Moreno_2021_emulate_target, Jackson, nguyen_clarifying_2021, vanderweelerobinson2014causalrace, Daniel2017}, the following sections distinguish three approaches to mediating mechanisms of social disparities. Interventional effects are integral to all three approaches.
Causal mediation is the focus of Approach 1. In this context, interventional (in)direct effects are discussed as alternative effect measures when natural (in)direct effects are not identified \citep[][]{VanderWeeleVansteeland_Robins_RIA2014, Daniel2017, Tylervanderweele_Tchetgen_2017mediation, Lin_Survival_2017, moreno2018understanding, Miles_2023}.
Approaches 2 and 3 are classified as pure interventional effects approaches \citep{nguyen_clarifying_2021}. Approach 2 evaluates the impact of hypothetical mediator manipulations on the exposure-induced disparity \citep{Moreno_2021_emulate_target}. In contrast to Approach 1, it does not aim to measure mediation.
Approach 3 focuses on the actual observed disparity and evaluates the impact of manipulations of explanatory factors - such as potential mediators, the focus of this article - on this disparity \citep{Jackson, vanderweelerobinson2014causalrace}. In contrast to Approach 1, it does not aim to measure mediation, and, in contrast to Approaches 1 and 2, it does not involve specifying causal effects of the exposure.

Approaches 1-3 are illustrated using the DAG in Figure \ref{Structural assumptions to illustrate Approaches 1-3}, which involves multiple interdependent mediators $M_1, \ldots, M_K$.
The indices assigned to the mediators represent a working order, which may not align with the true causal order. 
The focus will be on interventional effects, as proposed by \citet{Daniel2017}, \citet{Moreno_2021_emulate_target}, and \citet{Jackson}, that do not rely on the correct specification of the structural dependence among mediators. This acknowledges the practical difficulties associated with such specifications. The DAG in Figure \ref{Structural assumptions to illustrate Approaches 1-3} explicitly distinguishes between the mediators of interest and other upstream intermediates (i.e., those causally prior to the mediators), unlike the DAGs in previous studies \citep{Daniel2017, Moreno_2021_emulate_target, Jackson}. This distinction may be relevant in practical applications. An example is provided in Section \ref{Application}, where $\boldsymbol{L}$ involves sociodemographic covariates that may act as exposure-induced confounders.

For the effect definitions, the following additional notation is used: $\boldsymbol{M}$ denotes the set of the mediators $M_1, \ldots, M_K$, i.e., $\boldsymbol{M}=(M_1, \ldots, M_K)$. Furthermore, $\widetilde{\boldsymbol{M}}$ denotes the vector of random draws $\widetilde{M}_1, \ldots, \widetilde{M}_K$ from a user-specified joint intervention distribution. $\boldsymbol{I}$ denotes a subset of $\boldsymbol{M}$ such that $\boldsymbol{I} \subseteq \boldsymbol{M}$. The remaining mediators in $\boldsymbol{M}$ are summarized in the vector $\boldsymbol{R}$ such that $\boldsymbol{I} \cup \boldsymbol{R}=\boldsymbol{M}$. For any subset of mediators $\boldsymbol{I}$ (or $\boldsymbol{R}$), $\widetilde{\boldsymbol{I}}$ ($\widetilde{\boldsymbol{R}}$) denotes the vector of random draws from a user-specified joint intervention distribution of the variables in $\boldsymbol{I}$ ($\boldsymbol{R}$).
Unless otherwise stated, this article refers to average effects. Effects for the covariate strata in $\boldsymbol{C}$ can be obtained by conditioning on $\boldsymbol{C}$.
{Positivity of exposure conditions}\footnote{$P(A=a'|\boldsymbol{C})>0$ for all  $a'\in \{a,a^*\}$} {and of relevant mediator values in the support of the intervention distributions}\footnote{For example, if the intervention distribution of the mediators is specified to be $P_{\boldsymbol{M}_a|\boldsymbol{C}}(\boldsymbol{m}|\boldsymbol{C})$, positivity of relevant mediator values $\boldsymbol{m}$ refers to $P(\boldsymbol{M}=\boldsymbol{m}|A=a, \boldsymbol{C}=\boldsymbol{c}) >0$ for all $\boldsymbol{m}$ in the support of the intervention distribution where $P(A=a, \boldsymbol{C}=\boldsymbol{c})>0$ \citep[][]{NguyenSchmidOgburnStuart_2022}.}, as well as consistency{, which links counterfactual to observed values,}\footnote{$P(Y_{a \boldsymbol{m}} = Y)=1$ if $A=a$ and $\boldsymbol{M}=\boldsymbol{m}$, and $P(\boldsymbol{M}_{a} = \boldsymbol{M}) =  1$ if $A=a$.} are assumed throughout. As a reminder, the following sections focus on a binary exposure that takes values in $\{0,1\}$.

\subsection{Approach 1: Causal mediation analysis of the exposure-induced disparity}
\label{Approach1}
Approach 1, the causal mediation approach, aims to examine direct and indirect effects through which the exposure affects the outcome. The targeted estimands are typically natural effects and path-specific effects \citep{Pearl2001, robins1992identifiability, Avin2005, Shpitser2013}. However, these effects require strong assumptions that may not hold in scenarios with multiple mediators \citep{Vansteelandt_MultipleMediators2014}. For example, the natural indirect effect through a subset of mediators $\boldsymbol{I}$,  $\boldsymbol{I} \subseteq \boldsymbol{M}$, 
is generally not identified under the structural assumptions in the model in Figure \ref{Structural assumptions to illustrate Approaches 1-3}. This is due to exposure-induced confounding through $\boldsymbol{L}$ and potentially through other mediators in $\boldsymbol{R}$ that are upstream of $\boldsymbol{I}$. 

For settings with multiple interdependent mediators, \citet{Daniel2017} introduced a decomposition of the total effect, defined here as the exposure-induced disparity given by TE=$E[Y{_{1\boldsymbol{M}_1}}-Y{_{0\boldsymbol{M}_0}}]$, into interventional path-specific (in)direct effects. This approach does not require cross-world independence assumptions or assumptions about the structural dependence of the mediators. Consequently, it is also applicable when the causal ordering of mediators is unknown or when mediators share unmeasured common causes.
To illustrate this method and the challenges that arise when pursuing a conventional mediational interpretation, the following question will be addressed:\\

\textit{``What is the average indirect effect of $A$ on $Y$ through the mediators in $\boldsymbol{I}$, capturing all of the exposure effect that is mediated by $\boldsymbol{I}$, but not by causal descendants of $\boldsymbol{I}$ in the graph?''} \STEP{Question 1}\\

The method proposed by \citet{Daniel2017} suggests that \REF{Question 1} can be approached using an interventional indirect effect, defined by the contrast between two expected counterfactual outcomes under exposure resulting from two different stochastic mediator interventions. {The first intervention is defined by setting the values of the mediators in $\boldsymbol{I}$ to random draws from their counterfactual joint distribution under exposure, given $\boldsymbol{C}$, denoted by $\widetilde{\boldsymbol{I}}_{1|\boldsymbol{C}}$. Similarly, the values of the mediators in $\boldsymbol{R}$ are assigned random draws from their counterfactual joint distribution under exposure, given $\boldsymbol{C}$, denoted by $\widetilde{\boldsymbol{R}}_{1|\boldsymbol{C}}$. In summary, the mediators are set to a random draw from the distribution $P_{\boldsymbol{I}_{1|\boldsymbol{C}}}(\boldsymbol{i}|\boldsymbol{C}) \times P_{\boldsymbol{R}_{1|\boldsymbol{C}}}(\boldsymbol{r}|\boldsymbol{C})$. The second intervention is defined by setting the values of the mediators in $\boldsymbol{I}$ to random draws from their counterfactual joint distribution under control, given $\boldsymbol{C}$, denoted by $\widetilde{\boldsymbol{I}}{_{0|\boldsymbol{C}}}$. The distribution of the mediators in $\boldsymbol{R}$ is held constant at their counterfactual joint distribution under exposure, given $\boldsymbol{C}$. In summary, the mediators are set to random draws from the distribution $P_{\boldsymbol{I}_{0|\boldsymbol{C}}}(\boldsymbol{i}|\boldsymbol{C}) \times P_{\boldsymbol{R}_{1|\boldsymbol{C}}}(\boldsymbol{r}|\boldsymbol{C})$.  
These interventions define an interventional indirect effect through $\boldsymbol{I}$ as follows:
\begin{eqnarray}
    \label{IIE}
    \text{IIE}_{\boldsymbol{I}|\boldsymbol{C}} =  E[Y_{1 \boldsymbol{\widetilde{I}}_{1|\boldsymbol{C}}\boldsymbol{\widetilde{R}}_{1|\boldsymbol{C}}}-Y_{1 \boldsymbol{\widetilde{I}}_{0|\boldsymbol{C}} \boldsymbol{\widetilde{R}}_{1| \boldsymbol{C}}}]. 
\end{eqnarray}

Specifying the intervention distributions of both $\boldsymbol{I}$ and $\boldsymbol{R}$ marginally with respect to $\boldsymbol{L}$  
enables a decomposition of the TE into interventional effects, similar to that proposed by \citet{Daniel2017} (where $\boldsymbol{I}=M_2$, $\boldsymbol{R}=M_1$, and $\boldsymbol{L}=\emptyset$). This decomposition consists of $\text{IIE}_{\boldsymbol{I}|\boldsymbol{C}}$, as well as an interventional indirect effect through $\boldsymbol{R}$, an interventional indirect effect through the interdependence of the mediators, and an interventional effect of $A$ on $Y$ not through $\boldsymbol{I}$ and $\boldsymbol{R}$. \footnote{Specifically, these components are: 1)  $\text{IIE}_{\boldsymbol{I}|\boldsymbol{C}}$, 2) an  interventional indirect effect through $\boldsymbol{R}$, given by $\text{IIE}_{\boldsymbol{R}|\boldsymbol{C}}= E[Y_{1 \boldsymbol{\widetilde{I}}_{0|\boldsymbol{C}}\boldsymbol{\widetilde{R}}_{1|\boldsymbol{C}}}-Y_{1 \boldsymbol{\widetilde{I}}_{0|\boldsymbol{C}} \boldsymbol{\widetilde{R}}_{0| \boldsymbol{C}}}]$, 3) an interventional direct effect not through $\boldsymbol{I}$ and $\boldsymbol{R}$, given by $\text{IDE}_{\boldsymbol{M}|\boldsymbol{C}}= E[Y_{1 \boldsymbol{\widetilde{M}}_{0|\boldsymbol{C}}}-Y_{0 \boldsymbol{\widetilde{M}}_{0|\boldsymbol{C}}}]$, 4) the difference between the sum of these three interventional effects and the TE, which, in turn, constitutes a separate interventional indirect effect resulting from the effect of the exposure on the dependence between {mediators}, given $\boldsymbol{C}$ \citep[][]{Daniel2017, loh2021heterogeneous}.} 
For all possible subsets of $M_1,\ldots,M_K$ in $\boldsymbol{I}$ and $\boldsymbol{R}$, these components are nonparametrically identified under the following conditional independence assumptions \citep[][]{Daniel2017}, {interpreted as assumptions of no unmeasured confounding}: 
\begin{enumerate}
    \item [\ASSUMPTION{A1}] $Y_{a m_1 \ldots m_K} \independent A |\boldsymbol{C}$: no unmeasured confounding of the relationship between $A$ and $Y$ conditional on $\boldsymbol{C}$,
     \item [\ASSUMPTION{A2}] $Y_{a m_1 \ldots m_K} \independent (M_1, \ldots, M_K) | \{A=a, \boldsymbol{C}, \boldsymbol{L}\}$: no unmeasured confounding of the relationship between the mediators $M_1,\ldots, M_K$ and $Y$ conditional on $A=a$, $\boldsymbol{C}$, $\boldsymbol{L}$,
    \item [\ASSUMPTION{A3}] $(M_{1 a}, \ldots, M_{K a}) \independent A |\boldsymbol{C}$: no unmeasured confounding of the relationship  between $A$ and $M_1, \ldots, M_K$ conditional on $\boldsymbol{C}$.
\end{enumerate}

 If \ass{A1}, \ass{A2}, and \ass{A3} hold, $\text{IIE}_{\boldsymbol{I}|\boldsymbol{C}}$ is nonparametrically identified by a function of the observed data (see \citep[][]{Daniel2017} and Appendix, Section \ref{Identification}), given by:
\begin{eqnarray}
\label{Identification 1}
\sum_{\boldsymbol{c}}\sum_{\boldsymbol{l}} \sum_{m_1, \ldots, m_K} E[Y|A=1, \boldsymbol{c},\boldsymbol{l}, \boldsymbol{m}]  (P(\boldsymbol{i}|A=1, \boldsymbol{c})- P(\boldsymbol{i}|A=0, \boldsymbol{c}))   P(\boldsymbol{r}|A=1, \boldsymbol{c})  \\ 
P(\boldsymbol{l}|A=1,\boldsymbol{c})  P(\boldsymbol{c}). \nonumber
\end{eqnarray}

\ref{A1}, \ref{A2}, \ref{A3} hold under the structural assumptions in Figure \ref{Structural assumptions to illustrate Approaches 1-3}. They also hold when $\boldsymbol{L}$ and $\boldsymbol{M}$ share unmeasured common causes, or when $\boldsymbol{L}$ and $\boldsymbol{Y}$ share unmeasured common causes (see Figure \ref{Structural assumptions unmeasured common causes of intermediates} in the Appendix) \citep{Daniel2017, VanderWeeleVansteeland_Robins_RIA2014}.  However, these conditions do not ensure that $\text{IIE}_{\boldsymbol{I}|\boldsymbol{C}}$ satisfies the indirect effect measure criteria outlined by \citet{Miles_2023}.  These criteria consist of the sharp null, the sharper null, and the monotonicity criteria. The sharp(er) null criterion asserts that an indirect effect measure is equal to the null value when no individual indirect effect exists within the population of interest. The monotonicity criterion asserts that an indirect effect measure is not only capable of detecting the presence of an indirect effect, but is also able to correctly identify the direction of the mediated effect at least for some subjects. 
Based on a single-mediator model, \citet{Miles_2023} showed that the interventional indirect effect $\text{IIE}_{M|\boldsymbol{C}} =  E[Y_{1 \widetilde{M}_{1|\boldsymbol{C}}}-Y_{1 \widetilde{M}_{0|\boldsymbol{C}}}]$ can fail to satisfy these criteria. Scenarios in which this happens may be rare. For instance, the absence of overlap between the groups of individuals for whom the exposure affects the mediator and the groups of individuals for whom the mediator affects the outcome would be indicative of such an occurrence. Nonetheless, additional assumptions are necessary to exclude such scenarios and ensure a valid mediational interpretation of $\text{IIE}_{M|\boldsymbol{C}}$.
Either of the following two conditions is sufficient for $\text{IIE}_{M|\boldsymbol{C}}$ to satisfy the indirect effect measure criteria \citep{Miles_2023}: 
\begin{enumerate}
\label{Assumption A4-A5}
    \item [\ASSUMPTION{A4}] There are no exposure-induced confounders of the mediator and the outcome, or 
    \item [\ASSUMPTION{A5}] There is no mean interaction between exposure-induced confounders and the mediator {on the outcome on the additive scale}.
\end{enumerate}
Miles' findings suggest that the mediator can be either a single variable or a vector of variables. Thus, assumptions \ass{A4} and \ass{A5} can be mapped to $\text{IIE}_{\boldsymbol{I}|\boldsymbol{C}}$ with $\boldsymbol{I}=\boldsymbol{M}$. To assess the plausibility of \ref{A4} or \ref{A5} with respect to a proper mediator subset $\boldsymbol{I}$, assumptions must be made about which factors in $\boldsymbol{R}$ precede mediators in $\boldsymbol{I}$ and act as exposure-induced confounders in the $\boldsymbol{I} -Y$ relationship. The indirect effect measure criteria ensure that $\text{IIE}_{\boldsymbol{I}|\boldsymbol{C}}$ can be interpreted as an analogue of a natural path-specific indirect effect from $A$ to $Y$ via $\boldsymbol{I}$. This effect captures all pathways from $A$ to $\boldsymbol{I}$ (direct and via upstream intermediates), and from $\boldsymbol{I}$ to $Y$, excluding pathways passing through causal descendants of $\boldsymbol{I}$ in the graph. This interpretation holds regardless of whether the underlying causal structure among the mediators is known \citep{Daniel2017}.

\citet{tchetgen2014identification} showed that the natural indirect effect through a mediator is nonparametrically identified, even in the presence of an exposure-induced confounder, provided that assumption \ref{A5} holds. Therefore, under assumptions \ref{A4} or \ref{A5}, the interventional indirect effect analogue offers no practical advantage over the natural indirect effect, as both share the same identification formula \citep{Miles_2023}. If neither assumption holds, alternative indirect effects could be considered that meet the indirect effect measure criteria (see Appendix, Section \ref{AlternativeEffectmeasures}). However, these alternatives do not adequately address \ref{Question 1}.

To conclude, without additional assumptions beyond those required for its identification, $\text{IIE}_{\boldsymbol{I}|\boldsymbol{C}}$ may fail to adequately address \ref{Question 1}. These considerations with respect to \ref{Question 1} and $\text{IIE}_{\boldsymbol{I}|\boldsymbol{C}}$ demonstrate that mediation analysis in contexts with interdependent mediators is challenging, as it relies on heavy assumptions. Although interventional (in)direct effects were presented as a pragmatic approach to mediation in such complex settings \citep[][]{VanderWeeleVansteeland_Robins_RIA2014, Daniel2017}, the results of \citet{Miles_2023} raise doubts about whether this approach can truly facilitate mediation analyses. Nevertheless, even without a conventional mediational interpretation, interventional (in)direct effects may still provide valuable insights. For instance, $\text{IIE}_{\boldsymbol{I}|\boldsymbol{C}}$ provides a measure of the impact of differing counterfactual distributions of $\boldsymbol{I}$ under exposure and under control (given $\boldsymbol{C}$) on the outcome. It can be non-zero only if the exposure alters the distribution of $\boldsymbol{I}$, and this change in distribution affects the outcome \citep[]{Tylervanderweele_Tchetgen_2017mediation}. Such a measure may capture mediational concepts relevant from a population-level perspective \citep[][]{moreno2018understanding}. See \citet{Miles_2023} for alternative causal interpretations of interventional indirect effects.

\subsection{Approach 2: Evaluating the impact of manipulations of {mediating mechanisms} on the exposure-induced disparity}
\label{Approach2}
\citet[][]{Moreno_2021_emulate_target} described an interventional approach that evaluates the effects of mediator interventions, conceivable for hypothetical target trials, on the exposure-induced disparity. Based on this work, Approach 2 is defined, in which interventional effects are of intrinsic interest, rather than pragmatic alternatives. In contrast to Approach 1, Approach 2 does not aim to measure the causal mechanisms (i.e., the (in)direct effects) through which the exposure affects the outcome. Consequently, indirect effect measure criteria are not crucial for Approach 2.

As shown by \citet[][]{Moreno_2021_emulate_target}, various intervention strategies can be considered in a multi-mediator setting. For example, by evaluating interventions on each mediator separately, it is possible to determine which intervention would most reduce the exposure-induced disparity. A sequential intervention strategy allows the reduction in the exposure-induced disparity to be assessed by intervening on the mediators in a specified order. Generally, Approach 2 allows for the flexible specification of mediator interventions tailored to the scientific question at hand. As such, it is possible to fix mediators at certain values, resulting in degenerate mediator distributions. Consequently, controlled direct effects (CDE) and the portion eliminated ($\text{TE}-\text{CDE}$) \citep{robins1992identifiability, Pearl2001} can be categorized within Approach 2 \citep{Moreno_2021_emulate_target, NguyenSchmidOgburnStuart_2022}. CDEs are defined through interventions where the mediators under exposure and under control are set to constant values. For example, the average CDE when setting the mediators in $\boldsymbol{M}$ to the values $\boldsymbol{m}$ is defined by $\text{CDE}_{\boldsymbol{m}}=E[Y_{1\boldsymbol{m}}-Y_{0\boldsymbol{m}}]$. Interventional direct effects can be viewed as the average of controlled direct effects corresponding to various levels of the mediator, averaged according to a specified mediator distribution  \citep{VanderWeeleVansteeland_Robins_RIA2014}. They can be employed in Approach 2 to measure the exposure-induced disparity that would remain if the mediator distributions under exposure were equal to those under control.

The following estimand, which illustrates the approach, involves a similar intervention on the mediators as defined above for $\text{IIE}_{\boldsymbol{I}|\boldsymbol{C}}$ (equation \ref{IIE}). However, the scientific question differs. The aim is not to measure an indirect effect, but to evaluate the reduction in the exposure-induced disparity resulting from a specific mediator intervention. Specifically, the question is: \\

\textit{``What is the average reduction in the exposure-induced disparity (TE) achieved by setting the joint distribution of the mediators in $\boldsymbol{I}$ under exposure equal to what it would be in the counterfactual world under control, given $\boldsymbol{C}$, while keeping the joint distribution of the other mediators constant as it would be under exposure, given $\boldsymbol{C}$ and $\boldsymbol{L}$ under exposure?''}
\STEP{Question 2}\\

The mediator intervention corresponding to \REF{Question 2} involves setting the mediators in $\boldsymbol{I}$ and $\boldsymbol{R}$  under exposure to a random draw from the  distribution $P_{\boldsymbol{I}_{0|\boldsymbol{C}}}(\boldsymbol{i}|\boldsymbol{C}) \times P_{\boldsymbol{R}_{1|\boldsymbol{C}, \boldsymbol{L}_1}}(\boldsymbol{r}|\boldsymbol{C}, \boldsymbol{L}_1)$, where the counterfactual $\boldsymbol{L}_1$ refers to $\boldsymbol{L}$ under exposure. This intervention results in a shift of the joint distribution of $\boldsymbol{I}$ under exposure to match the counterfactual distribution under control, given $\boldsymbol{C}$,  thus severing the dependence of $\boldsymbol{I}$ on $\boldsymbol{L}$ and $\boldsymbol{R}$. Meanwhile, the joint distribution of $\boldsymbol{R}$ remains unchanged under exposure, preserving the dependence among $\boldsymbol{L}$ and $\boldsymbol{R}$. 
The reduction in the TE resulting from this intervention is an interventional effect that refers to the change in the counterfactual outcome under exposure. Formally, it is given by: 
\begin{eqnarray}
\label{IE_I}
  \text{IE}_{\boldsymbol{I}|\boldsymbol{C}} =  E[Y_{1 \boldsymbol{M}_1}-Y_{1 \boldsymbol{\widetilde{I}}_{0|\boldsymbol{C}} \boldsymbol{\widetilde{R}}_{1|\boldsymbol{C}, \boldsymbol{L}_1}}].
\end{eqnarray}

 The residual exposure-induced disparity, given the world under control remains unchanged, is given by $\text{RE}_{\boldsymbol{I}|\boldsymbol{C}}=E[Y_{1 \boldsymbol{\widetilde{I}}_{0|\boldsymbol{C}} \boldsymbol{\widetilde{R}}_{1|\boldsymbol{C}, \boldsymbol{L}_1}}-Y_{0 \boldsymbol{M}_0}]$. To identify $\text{IE}_{\boldsymbol{I}|\boldsymbol{C}}$ and $\text{RE}_{\boldsymbol{I}|\boldsymbol{C}}$, assumptions \ref{A1}-\ref{A3} are required  \citep[][]{Moreno_2021_emulate_target, NguyenSchmidOgburnStuart_2022}.\footnote{As stated by \citet{nguyen_clarifying_2021}, conditioning on $L_a$ requires consistency of $L_a$ and the conditional independence assumption $(L_a, M_{1a}, \ldots, M_{Ka}) \independent A|\boldsymbol{C}$, $a \in \{0,1\}$. This independence is implied by assumption \ref{A3}: $(M_{1a}, \dots, M_{Ka}) \independent A | \boldsymbol{C}$, since $\boldsymbol{L}_a$ is a cause of $\boldsymbol{M}_a$. Therefore, $\boldsymbol{C}$ must also include confounders of the $A - \boldsymbol{L}$ relationship.}
\ref{A1} and \ref{A2} ensure that the effects of exposure and mediator interventions on the outcome are identified. \ref{A3} ensures that the counterfactual mediator distributions under different exposure conditions (given $\boldsymbol{C}$) are identified. If \ref{A1}-\ref{A3} hold, $\text{IE}_{\boldsymbol{I}|\boldsymbol{C}}$ is nonparametrically identified for each possible mediator subset of $M_1, \ldots, M_K$ in $\boldsymbol{I}$ by (see Appendix, Section \ref{Identification}, and \citet[][]{Moreno_2021_emulate_target}, where $\boldsymbol{I}=M_k (k=1, \ldots, K)$ and $\boldsymbol{L}=\emptyset$)

  \begin{align}
 \label{IE Form}
  \sum_{\boldsymbol{c}}\sum_{\boldsymbol{l}} \sum_{m_1, \ldots, m_K}  E[Y|A=1, \boldsymbol{c},\boldsymbol{l}, \boldsymbol{m}]  (P(\boldsymbol{m}|A=1, \boldsymbol{l}, \boldsymbol{c}) - 
  P(\boldsymbol{i}|A=0, \boldsymbol{c}) 
   P(\boldsymbol{r}|A=1, \boldsymbol{l}, \boldsymbol{c}))   \\ \nonumber
 P(\boldsymbol{l}|A=1,\boldsymbol{c})  P(\boldsymbol{c}). 
 \end{align}

\subsection{Approach 3:  Evaluating the impact of manipulations of explanatory mechanisms on the actual observed disparity}
\label{Approach3}
\citet{vanderweelerobinson2014causalrace} and \citet{Jackson} outlined a framework for decomposing actual observed disparities, referred to as Approach 3. This approach evaluates the effects of altering the distributions of explanatory variables, such as potential mediators or variables that lie on a backdoor path from the exposure to the outcome (e.g., common causes of the exposure and the outcome), on the exposure-outcome association. Such evaluations help identify intervention targets aimed at reducing actual observed social disparities \citep[]{Jackson}. Unlike Approaches 1 and 2, Approach 3 does not aim to identify causal effects of the exposure. This makes it a viable option in cases where there are unmeasured confounders of the exposure-outcome or exposure-mediator relationships, or when interventions on the exposure would be ill-defined (e.g., because the exposure is non-manipulable). Furthermore, the method is adaptable to multiple-mediator settings and, like Approach 2, offers flexibility in specifying intervention strategies and distributions.\\
The following question, which serves to illustrate the approach, involves interventions on hypothesized mediators that may resemble those defined above for \(\text{IE}_{\boldsymbol{I}|\boldsymbol{C}}\) (equation \ref{IE_I}). The key difference lies in the nature of the intervention distributions. Unlike before, the question refers to an observed intervention distribution rather than a counterfactual one. Specifically, the question is:\\

\textit{``What is the average reduction in the actual observed outcome disparity, $E[Y|A=1]-E[Y|A=0]$, achieved by setting the joint distribution of the mediators in $\boldsymbol{I}$ among the exposed group equal to that among the unexposed group, given $\boldsymbol{C}$, while keeping the joint distribution of the other mediators constant, given $\boldsymbol{C}$ and $\boldsymbol{L}$?''}
\STEP{Question 3}
\\

The intervention associated with \REF{Question 3} sets the mediators in $\boldsymbol{I}$ and $\boldsymbol{R}$ among the exposed to a random draw from the distribution {$P_{\boldsymbol{I}|A=0,\boldsymbol{C}}(\boldsymbol{i}|A=0, \boldsymbol{C})$} {$\times P_{\boldsymbol{R}|A=1,\boldsymbol{C}, \boldsymbol{L}}(\boldsymbol{r}|A=1, \boldsymbol{C}, \boldsymbol{L})$}.
As a consequence, the joint distribution of $\boldsymbol{I}$ among the exposed is aligned with that among the unexposed, given $\boldsymbol{C}$, thus removing the dependence of $\boldsymbol{I}$ on $\boldsymbol{L}$ and $\boldsymbol{R}$. The resulting reduction in the observed disparity is a change in the mean of $Y$ among the exposed and is formally given by

\begin{eqnarray}
\label{IEobsI}
  \text{IE}_{\boldsymbol{I}|\boldsymbol{C}(obs)} &=&  E[Y|A=1]- E[Y_{\widetilde{\boldsymbol{I}}|A=0, \boldsymbol{C} \,{\widetilde{\boldsymbol{R}}|A=1,\boldsymbol{C}, \boldsymbol{L}}}|A=1].
\end{eqnarray}

 The residual disparity, given the world among the unexposed remains unchanged, is given by $\text{RE}_{\boldsymbol{I}|\boldsymbol{C}(obs)}=E[Y_{\widetilde{\boldsymbol{I}}|A=0, \boldsymbol{C} \,{\widetilde{\boldsymbol{R}}|A=1,\boldsymbol{C}, \boldsymbol{L}}}|A=1]- E[Y|A=0]$, which is a non-causal estimand. 
The counterfactual $E[Y_{\widetilde{\boldsymbol{I}}|A=0, \boldsymbol{C} \,{\widetilde{\boldsymbol{R}}|A=1,\boldsymbol{C}, \boldsymbol{L}}}|A=1]$ is identified if there is no unmeasured confounding of the relationship between $M_1,\ldots, M_K$ and $Y$ conditional on $A=1$, $\boldsymbol{C}$, $\boldsymbol{L}$. In particular, $\text{IE}_{\boldsymbol{I}|\boldsymbol{C}(obs)}$ is then nonparametrically identified by (see Appendix, Section \ref{Identification} and \citep[][]{Jackson} for a related estimand)

\begin{eqnarray}
E[Y|A=1]- \sum_{\boldsymbol{c}}\sum_{\boldsymbol{l}} \sum_{m_1, \ldots, m_K} E[Y|A=1, \boldsymbol{c},\boldsymbol{l}, \boldsymbol{m}]  P(\boldsymbol{i}|A=0, \boldsymbol{c})  
 P(\boldsymbol{r}|A=1, \boldsymbol{l}, \boldsymbol{c})  \\ P(\boldsymbol{l}|A=1, \boldsymbol{c}) P(\boldsymbol{c}|A=1). \nonumber
\end{eqnarray}

Based on the structural assumptions in Figure \ref{Structural assumptions to illustrate Approaches 1-3}, $\text{IE}_{\boldsymbol{I}|\boldsymbol{C}(obs)}$ is nonparametrically identified. Notably, it would also be identified under weaker assumptions, such as when $A$ and $Y$, or $A$ and $\boldsymbol{M}$, share unmeasured common causes (see Figure \ref{Structural assumptions to illustrate Approaches3} in the Appendix).

\citet{Jackson} linked their proposed interventional framework to the twofold OB decomposition method. Building on their findings, the next section explores the use of the twofold OB decomposition for linear outcome models within Approaches 1-3.

\section{The twofold Oaxaca-Blinder decomposition {for linear outcome models}}
\label{Section Oaxaca}

This section addresses the applicability of the twofold OB decomposition for linear outcome models as an estimator within the three approaches described in Section \ref{Section: Interventional effects general}. It is guided by pre-specified nonparametric causal estimands, structural assumptions, and modeling assumptions. For illustrative purposes, the focus will be on settings with a single mediator and a single exposure-induced confounder, as illustrated in the models presented in the Figures \ref{Graph2} and \ref{Graph1}.

\subsection{Specification and identification of causal estimands}
Consider the following scientific questions and the corresponding causal estimands:\\
\begin{enumerate}
\item \textit{What is the average reduction in the actual observed outcome disparity achieved by setting $M$ among the exposed equal to its marginal mean among the unexposed?} \STEP{Question 4} \\

According to the interventional effects classification in Section \ref{Section: Interventional effects general}, \REF{Question 4} represents a specific causal estimand within Approach 3, given by 

\begin{eqnarray}
\label{IEobsM}
\text{IE}_{M(obs)}=E[Y-Y_{E[M|A=0]}|A=1].
\end{eqnarray}

The remaining exposure-outcome association is given by $\text{RE}_{M(obs)}=E[Y_{E[M|A=0]}|A=1]-E[Y|A=0]$.
If  the conditional independence assumption $Y_{m}\independent M |A=1, C, L$ holds as in the models in Figures \ref{Graph2} and \ref{Graph1}, $E[Y_{E[M|A=0]}|A=1]$ is nonparametrically identified by (see \citep[][]{Jackson} for a related estimand)

\begin{eqnarray}
\label{OB NONPARA}
   \sum_{\boldsymbol{c}}\sum_{\boldsymbol{l}} E[Y|A=1, M=E[M|A=0], l, c]  P(l|A=1, c)   P(c|A=1).
\end{eqnarray}

\item \textit{What is the average reduction in the exposure-induced disparity achieved by setting $M$ under exposure equal to the  counterfactual expectation of $M$ under control?} \STEP{Question 5} \\

\REF{Question 5} represents a specific causal estimand within Approach 2, given by 
\begin{eqnarray}
\label{IEM}
 \text{IE}_{M}=E[Y_{1}-Y_{1 E[M_{0}]}]. 
\end{eqnarray}
The remaining exposure-induced disparity is defined by $\text{RE}_{M}=E[Y_{1 E[M_{0}]}-Y_{0}]$.
 $\text{IE}_{M}$ and $\text{RE}_{M}$ are nonparametrically identified under the structural assumptions of the DAGs in Figures \ref{Graph2} and \ref{Graph1}  (see \citep[][]{NguyenSchmidOgburnStuart_2022} for a similar context). Specifically, under those in Figure \ref{Graph1}, which imply $Y_a\independent A$, $Y_{am}\independent M|A=a, C, L$, and $M_a \independent A$,  $E[Y_{1 E[M_{0}]}]$ is nonparametrically identified by the formula in equation \ref{OB NONPARA}.\footnote{Given that $A$ is assumed to be independent of $C$ (see DAG in Figure \ref{Graph1}), it follows that $P(c|A=1)=P(c|A=0)=P(c)$.}

\item \textit{What is the {average} indirect effect of $A$ on $Y$ through $M${, capturing all pathways from the exposure to the outcome through $M$?}} \STEP{Question 6} \\

\REF{Question 6} represents a causal estimand within Approach 1, defined here as  the natural indirect effect through $M$ given by
\begin{eqnarray}
\label{NIE_M}
\text{NIE}_{M}=E[Y_{1M_{1}}-Y_{1M_{0}}].
\end{eqnarray}
In general, however, $\text{NIE}_{M}$ is not identified under the structural assumptions of the DAGs in Figures \ref{Graph2} and \ref{Graph1} , as they involve an exposure-induced confounder $L$. As an alternative, one may consider the interventional indirect effect, which is defined by setting the marginal distribution of $M$ under exposure equal to that under control. Formally, this interventional indirect effect is given by
\begin{eqnarray}
\label{IIEM}
 \text{IIE}_{M}=E[Y_{1\widetilde{M}_{1}}-Y_{1 \widetilde{M}_{0}}],  
\end{eqnarray} 
with $\widetilde{M}_{a}$, $a\in \{0,1\}$, denoting a random draw from $P_{M_a}(m) = E_C[P_{M_a|C}(m|C)]$ \citep{NguyenSchmidOgburnStuart_2022}. The corresponding interventional effect not through $M$ is given by  $\text{IDE}_{M}=E[Y_{1\widetilde{M}_{0}}-Y_{0 \widetilde{M}_{0}}]$. 

$\text{IIE}_{M}$ and $\text{IDE}_{M}$ are nonparametrically identified under the structural assumptions of the DAGs in Figures \ref{Graph2} and \ref{Graph1}.
Specifically, under the DAG in Figure \ref{Graph1}, $\text{IIE}_{M}$ is identified by (see identification results in \citep{NguyenSchmidOgburnStuart_2022}, also with respect to the DAG in Figure \ref{Graph2})
\begin{eqnarray}
\label{OB IIE NONPARA}
  \sum_{\boldsymbol{c}}\sum_{\boldsymbol{l}}\sum_{{m}} E[Y | A=1, m, l, c]  (P(m|A=1)-P(m|A=0))P(l|A=1, c)   P(c).
\end{eqnarray}
 If $L=\emptyset$ (\ref{A4}), or if there is no mean interaction between $L$ and $M$  on $Y$ on the additive scale (\ref{A5}) \citep{Miles_2023}, $\text{IIE}_{M}$ satisfies the indirect effect measure criteria, and thus serves as an indirect effect analogue to $\text{NIE}_{M}$.
 
As a side note, \citet{VanWeele_Steel2009} demonstrated with respect to natural (in)direct effects that, if the outcome is linear in the mediator, it suffices to correctly specify the mediator's expectation rather than its entire distribution (for binary outcomes see \citep{VanderWeele_Vansteelandt_Odds_2010, TchetgenTchetgen_Odds_2014}). Extending these results to interventional effects suggests that, under linearity of the outcome in the mediator, $\text{IE}_{M(obs)}$ (equation \ref{IEobsM}) can be interpreted as the change in the expected outcome among the exposed when the distribution of $M$ is aligned with its marginal distribution among the unexposed. Furthermore, $\text{IE}_{M}$ (equation \ref{IEM}) can be interpreted as the change in the counterfactual outcome under exposure when aligning the distribution of $M$ under exposure with its marginal distribution under control.

\end{enumerate}
\subsection{Estimation using a variant of the twofold Oaxaca-Blinder decomposition {for linear outcome models}}
Consider the following linear outcome models for the exposed group (A=1) and the unexposed group (A=0):

\begin{eqnarray}
\label{ModelOB Exposure}
E[Y|A=1, M=m, L=l, C=c] &=& \alpha_0 + \alpha_1m + \alpha_2 l + \alpha_3 ml +\alpha_4c
\end{eqnarray}
\begin{eqnarray}
\label{ModelOB Control}
E[Y|A=0, M=m, L=l, C=c] &=& \omega_0 + \omega_1 m + \omega_2 l + \omega_3 m l + \omega_4 c .
\end{eqnarray}

The marginal mean of $Y$ in the exposure group is obtained by:
\begin{flalign}
E[Y|A=1] =
\alpha_0+\alpha_1E[M|A=1] + \alpha_2E[L|A=1] + \alpha_3E[ML|A=1] + \\ \nonumber
\alpha_4  E[C|A=1] .
\end{flalign}
Equally, the marginal mean of $Y$ in the unexposed group is obtained by:
\begin{flalign}
E[Y|A=0] =  
  \omega_0+\omega_1E[M|A=0] + \omega_2E[L|A=0] + \omega_3 E[ML|A=0] + \\ \nonumber
 \omega_4E[C|A=0] .
\end{flalign}

Using the standard twofold OB decomposition technique \citep{Oaxaca1973, Blinder1973}, the marginal disparity $E[Y|A=1]-E[Y|A=0]$ can be decomposed into two components:\\
\begin{equation}
\label{OB classic Decomposition}
    \begin{aligned}
 &E[Y|A=1]-E[Y|A=0] =   
    \end{aligned}
\end{equation}
\begin{equation}
\left.
\begin{aligned}
\label{OB_explained}
&\alpha_1 (E[M|A=1] - E[M|A=0]) + \alpha_2(E[L|A=1]-E[L|A=0]) +\\\nonumber
&\alpha_3(E[ML|A=1]-E[ML|A=0]) + 
\alpha_4  (E[C|A=1]-E[C|A=0]) + \nonumber
  \end{aligned}
\quad\right\}\quad \text{explained}
\end{equation}

\begin{equation}
\left.
\begin{aligned}
\label{OB_unexplained}
&(\alpha_0-\omega_0) +\big(\alpha_1 - \omega_1\big)E[M|A=0] +\big(\alpha_2 - \omega_2\big) E[L|A=0] +  \\\nonumber
&\big(\alpha_3 - \omega_3\big)E[ML|A=0] + \big(\alpha_4 - \omega_4\big)E[C|A=0],
 \end{aligned}
\quad\right\}\quad \text{unexplained}
\end{equation}
where the explained part refers to the part associated with (weighted) differences in the marginal means of the explanatory variables, and the unexplained part refers to the part associated with (weighted) differences in the coefficients.  These two components can be further decomposed into the contributions of individual covariates, with the interaction between $M$ and $L$ being treated as a distinct covariate.
When the focus is specifically on group differences in $M$, while taking into account the interaction between $M$ and $L$, the following two components may be of interest:  
\begin{align}
    \text{OB}_{M}&=&E[Y|A=1] - E[Y|A=1, M=E[M|A=0], L=E[L|A=1], C=E[C|A=1]], 
\end{align}
the component associated with differing means of $M$,  while accounting for the interaction of $M$ with $L$ and holding $L$ and $C$ fixed at their expected values in the exposed group, and 
\begin{align}
    \text{OB}_{RE}&=&E\left[Y|A=1, M=E\left[M|A=0\right], L=E\left[L|A=1\right], C=E\left[C|A=1\right]\right]-
    E[Y|A=0],
\end{align}
the component not associated with differing means of $M$, but with differences in coefficients and differences in the expected values of $L$ and $C$, as well as the interaction between $M$ and $L$. Together, $\text{OB}_{M}$ and $\text{OB}_{RE}$ add up to the marginal disparity. This twofold decomposition, which is considered here as a variation of the twofold OB method, can be performed as follows:
\begin{equation}
\label{OB Decomposition}
    \begin{aligned}
 &E[Y|A=1]-E[Y|A=0] =   
    \end{aligned}
\end{equation}
\begin{equation}
\left.
\begin{aligned}
\label{OB_M}
&\alpha_1 \big(E[M|A=1] -
 E[M|A=0]\big) + \\ \nonumber
 &\alpha_3 \big(E[ML|A=1] - 
 E[M|A=0] E[L|A=1]\big) + 
  \end{aligned}
\quad\right\}\quad \text{OB}_{M}
\end{equation}

\begin{equation}
\left.
\begin{aligned}
\label{OB_Res}
&\alpha_0-\omega_0 +\big(\alpha_1 - \omega_1\big)E[M|A=0] +\alpha_2 E[L|A=1]- \omega_2E[L|A=0] + \\ \nonumber
& \alpha_3 \big(E[M|A=0] E[L|A=1]\big)-\omega_3\big(E[ML|A=0]\big) +\\ \nonumber
& \alpha_4E[C|A=1]- \omega_4E[C|A=0].
 \end{aligned}
\quad\right\}\quad \text{OB}_{RE}
\end{equation}
\noindent

Without further assumptions, $\text{OB}_{M}$ and $\text{OB}_{RE}$ are statistical quantities that measure differences between conditional expected means.  Table \ref{OB Causal Estimands} summarizes the nonparametric and parametric assumptions, which are required to use $\text{OB}_{M}$ to estimate the causal estimands $\text{IE}_{M(obs)}$ (equation \ref{IEobsM}), $\text{IE}_{M}$ (equation \ref{IEM}) and $\text{IIE}_{M}=\text{NIE}_{M}$ (equations \ref{NIE_M} and \ref{IIEM}). They are cumulative, meaning each row's assumption assumes the previous rows' assumptions also hold. For completeness, the estimands corresponding to $\text{OB}_{RE}$ are also listed.

\begin{table}[H]
\begin{tabular}{|l|l|l|}

\hline
\textbf{Assumptions} & $\text{\textbf{OB}}_{M}$ & $\text{\textbf{OB}}_{RE}$\\
\hline
Consistency  & $\text{IE}_{M(obs)}=$&$\text{RE}_{M(obs)}=$  \\
Positivity &$E[Y-Y_{E[M|A=0]}|A=1]$&$E[Y_{E[M|A=0]}|A=1]-E[Y|A=0]$  \\
$Y_{m}\independent M|A=a, C, L$& & (non causal)\\
Linear outcome model && \\
Correct model specification  & & \\

    \hline
     $Y_{am}\independent M|A=a, C, L$  &$\text{IE}_{M}=E[Y_{1}-Y_{1E[M_{0}]}]$  &$\text{RE}_{M}=E[Y_{1E[M_{0}]}-Y_{0}]$ \\
    Exogeneity of the exposure: &&\\
    
    $Y_{am}\independent A$  &  &  \\
  
     $M_a\independent A$ &&\\ 
      \hline
  \ref{A4} ($L=\emptyset$) \citep{Huber2015, Miles_2023}, or& $\text{IIE}_{M}= E[Y_{1\widetilde{M}_{1}}-Y_{1 \widetilde{M}_{0}}]$ & $\text{IDE}_{M}=E[Y_{1\widetilde{M}_{0}}-Y_{0 \widetilde{M}_{0}}]=$ \\ 
  \ref{A5}  (no mean interaction& $=\text{NIE}_{M}$\footnotemark   & $\text{NDE}_{M}$ \\
  between $L$ and $M$)\citep{Miles_2023} & &\\
  \hline
  
\end{tabular}
\caption{Cumulative assumptions required for using $\text{OB}_{M}$ to estimate $\text{IE}_{M(obs)}$, $\text{IE}_{M}$, and $\text{IIE}_{M}$=NIE$_{M}$, with $a\in \{0,1\}$. For completeness, estimands corresponding  to $\text{OB}_{RE}$ are also listed.}
\label{OB Causal Estimands}
\end{table}
\footnotetext{It should be noted that the cross-world independence assumption is still required for natural effects.  However, under assumption \ref{A4} or assumption \ref{A5}, the identification formula for $\text{NIE}_{M}$ ($\text{NDE}_{M}$) aligns with that for $\text{IIE}_{M}$ ($\text{IDE}_{M}$), as outlined by \citet{Miles_2023}.}

As detailed in Table \ref{OB Causal Estimands}, the implementation of the proposed decomposition within Approach 3 requires the fewest assumptions. Conversely, its application within Approaches 1 and 2 entails considerably stronger conditions, including the exogeneity of the exposure. \citet[][]{Jackson} suggested applying the OB method to the strata of baseline covariates (i.e., conducting conditional OB decompositions), as this technique can have important implications for causal inference. For example, when the exposure shares a common cause $C$ with the mediator or the outcome, as in Figure \ref{Graph2}, $(Y_a, M_a)\independent A$ does not hold, but $(Y_a, M_a)\independent A|C=c$ does. In conclusion, given the varying degrees of strength of the assumptions involved, it can be stated that OB decompositions (and their variations) are most applicable to Approach 3.

Correct model specification remains crucial for unbiased estimation in all approaches. In settings with complex relationships, such as various (possibly higher-order) interactions or nonlinearities, standard OB decompositions (or their variations) may prove inadequate for estimating the targeted estimand or may be impractical.  In such cases, other estimation methods that allow for flexible modeling, such as g-computation \citep{Daniel2017}, or semiparametric and nonparametric multiply-robust methods \citep{tchetgen2012semiparametric, RubinsteinBransonKennedy+2023}, may be more appropriate. Section \ref{MC g-computation} of this article outlines the g-computation approach employed in the interventional effects analysis of the gender pay gap.

\section{Application of the interventional effects framework to the gender pay gap}
\label{Application}
\subsection{Preliminary considerations on the choice of approach}
In studying the gender pay gap, Approach 1 would be of interest in examining the direct and indirect effects of gender on wages, while Approach 2 would be used to evaluate the reduction of gender-induced wage disparity through relevant mediator interventions. Both approaches would entail the construction of causal effects pertaining to gender, implying the notion of hypothetical manipulations of gender. However, these concepts become somewhat ambiguous without a precise definition of gender or of which aspect of gender is in view (e.g., social roles, norms, perceptions, behaviors, or physical attributes). 
There is a debate over whether valid causal inference is possible with respect to ill-defined or non-manipulable exposures such as gender or race \citep[][]{holland1986statistics, vanderweele_hernan_2012causal, Glymour_2014_comment, Glymour_Spiegelman_2017, Pearl_2018_NonManipulable}. Some authors suggest shifting the focus from actual traits to perceived traits, which facilitates experimental manipulation \citep[][]{fienberg2003discussion, kaufman_epidemiologic_2008, Greiner_Rubin_2011_Perceived_Immutable_Characteristics}. For example, researchers can obscure the visible attributes of an applicant's gender by employing methods such as placing a screen between candidates and decision-makers \citep{goldin2000orchestrating}.
The illustrative model presented in Figure \ref{DAG_applied_PERCEPTION} for the effect of gender on wage involves gender perception. The model posits that gender affects wage through gender perception, suggesting that wage decisions may be influenced by whether an individual is perceived as a man or a woman. The meaning of ``gender perception'' will not be further elaborated upon here; instead, the focus will be on analytical techniques when assuming a model like that in Figure \ref{DAG_applied_PERCEPTION}. In this model, the pathway from gender to wage via gender perception could be addressed using the concept of controlled direct effects \citep{Pearl2001}. This approach would set all mediators - except gender perception - at equal levels for both women and men. Consequently, any remaining wage disparity could be attributed to a direct effect of gender through gender perception. However, this method relies on challenging conditions. Apart from conditional independence assumptions \citep{Tyler_book}, the adjusted set of mediators must be exhaustive to accurately separate the Gender $\rightarrow$ Gender Perception $\rightarrow$ Wage pathway from other pathways. 

Approach 3 circumvents the pitfalls of imprecise causal attributions to gender or gender categories as measured in a survey. The approach can be employed to evaluate the contribution of differences between men and women in the distributions of wage-relevant factors to the actual observed pay gap, without the need to define causal effects of gender. By simulating relevant hypothetical scenarios (e.g., where mediators are evenly distributed across men and women), intervention targets can be identified to reduce the gap.
 Approach 3 is illustrated in the following sections through an empirical analysis of the 2017 gender pay gap in Western Germany. Its comprehensive nature is shown by defining different mediator intervention distributions and directing interventions at women alone (Aims 1a-b), as well as at both women and men (Aims 2a-b), thereby addressing considerations regarding hypothetical changes in the working lives of both groups. For a comparison of Approach 3 with Approaches 1 and 2 with respect to the empirical analysis, see Section \ref{Contextualizing Aims} of the Appendix. This section of the Appendix summarizes key conditions under which the estimated effects represent specific quantities within Approaches 1-3, suggesting possible interpretations.

\begin{figure}[H]
\centering
 \scalebox{0.9}{
\begin{tikzpicture}[x=10in,y=6in]
\node (A) at (-0.05,-0.300) {Gender};
\node (J) at (0.15,-0.250) {Job characteristics};
\node (P) at (0.15,-0.350) {Gender perception};
\node (H) at (0.15,-0.140) {Human capital};

\node (Y) at (0.35,-0.300) {Wage};

\draw [->] (A) edge (H);
\draw [->] (A) edge (J);
\draw [->] (A) edge (P);

\draw [->] (P) edge (Y);

\draw [->] (J) edge (Y);
\draw [->] (H) edge (Y);

\draw [loosely dotted, very thick]  (H) edge (J);
\draw [loosely dotted, very thick]  (J) edge (P);

\draw [loosely dotted, very thick] (H) to [out=-5,in=0, looseness=0.7] (P);

\end{tikzpicture}
}

\caption{Conceptual overview of the factors assumed to mediate the gender pay gap, including ``Gender perception''. Dotted lines indicate that there may be dependencies among these factors. This figure is illustrative and does not represent a formal DAG.}
\label{DAG_applied_PERCEPTION}
\end{figure}
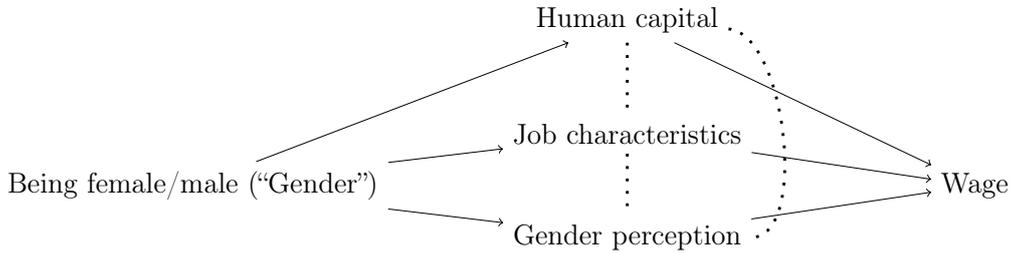

\subsection{Data and methods}
\label{Applied Analysis: Data and methods}
This section illustrates a cross-sectional application of Approach 3 to the gender pay gap in the labor market of Western Germany, using data from the 2017 SOEP, a large-scale, representative longitudinal survey of private households \citep{SOEP, goebel2019German}. The analysis examined how, under certain assumptions, hypothetical manipulations of wage-relevant factors could have narrowed the observed gender pay gap between women and men. The focus was on ten potential mediators, including indicators of education, labor market experience, and characteristics of employment. These variables, which are presented with descriptive statistics in Table \ref{8Mediators},  include concepts that have been considered as relevant explanatory factors in gender pay gap analyses \citep[][]{Golding_2014, Blau_Kahn_2017, Magnusson_Prestige_GPG, kunze2018gender}.
In general, categorical mediators with more than two categories were summarized into binary variables. Continuous mediators with highly skewed distributions were dichotomized to ease interpretation and to ensure that the parametric assumptions for estimation were met. The covariate set included age in years ($C$), an indicator for the presence of at least one child under the age of 18 in the household ($L_1$), and an indicator for a direct migration background ($L_2$) (see Table \ref{8Mediators}). Based on prior research, these covariates were considered relevant to various labor market outcomes \citep{becker1975ageearnings, Goldin_2006_Quiet_Revolution, Algan_2010_MigrationGap, Kleven_2019_ChildPenalties, Correll_Motherhood_Penalty_2007_AJS, Hipp_Children_Hiring_2019}. 
Gross hourly wages were calculated from the current gross labor income and average weekly working hours at the time of the survey, as generated by the SOEP \citep{PGEN2017v34}. This calculation follows the method proposed by \citet{brenke2013gesetzlicher}, which takes into account both compensated and uncompensated overtime, as well as actual and contracted hours. Due to the skewed distribution of gross hourly wages, the natural logarithm of gross hourly wages was used as the primary outcome. Assumptions about the data-generating process are illustrated in Figure \ref{DAG_applied} in the Appendix. 

\begin{table}[H]
    \centering
    \begin{adjustbox}{width=\textwidth}
    
            \begin{tabular}{p{0.06\textwidth}|p{0.6\textwidth}|p{0.175\textwidth}|p{0.175\textwidth}|p{0.175\textwidth}}\hline
                \textbf{$\boldsymbol{M}$} & \textbf{Mediators listed according to the working order}  & Total (11,924) & Women (6,182) & Men (5,742) \\
                & & Mean (IQR) & Mean (IQR) & Mean (IQR)\\
                \hline
               
                $M_1$ & ``$\geq$ 12 years of education'' (yes=1) & 0.50 & 0.51 & 0.48\\
                $M_2$ & College degree or higher (``$\geq$ College degree'', yes=1)  &0.25 & 0.23  & 0.27 \\
                $M_3$ & Current job requiring at least a college degree (``$\geq$ College degree job'', yes=1)  & 0.26 & 0.22  & 0.30\\
                $M_4$ & $\geq$ the median of 6.7 years of employment with company (``$\geq$ 6.7 years in company'', yes=1) & 0.50 & 0.46 & 0.54 \\
                $M_5$ & Job in on of the following industries: services, trade, banking, insurance (``Female-dominated industry'', yes=1)\tablefootnote{$M_5$ is based on the 1 Digit Industry Code of Individual as generated by the SOEP \citep{Grabka2024_pequiv}. 
                }  &0.62 & 0.78 & 0.45 \\
                $M_6$ & Leading function with managerial responsibility: senior civil servants, white-collar employees with highly qualified  work or comprehensive management responsibility  (``Leading position'', yes=1) &0.20   & 0.12 & 0.27 \\
                $M_7$ & ``Full-time employment'' (yes=1) & 0.64 & 0.40 & 0.90 \\
                $M_8$ & ``Flexible working hours'' (yes=1) &  0.41 &0.37 & 0.49  \\ 
                $M_9$ &  Work experience in years\tablefootnote{Full-time and part-time work experience in years was generated by the SOEP using combined monthly and annual employment data. From these measures, a measure for work experience was created, with one year of part-time experience counted as 0.5 years of full-time experience.} relative to age (``Work experience'') & 0.36 (0.21; 0.51) & 0.29 (0.17; 0.40)&  0.42  (0.30; 0.57)\\
                
                $M_{10}$ & Standard International Occupational Prestige Scale, based on the International Standard Classification of  Occupations 1988 (``Job prestige (SIOPS)'') & 44.56  (33; 53)  & 44.08 (33; 53)  & 45.08 (34; 56) \\
               
                \hline
                $C$, $\boldsymbol{L}$ & \textbf{Covariates } &  &  &  \\
              
                \hline
                $C$ & Age in years & 44.33 (36; 53)  & 44.21 (36; 53)   & 44.47 (36; 53)  \\ 
                $L_1$ & Child in household  (yes=1) & 0.48 & 0.47 & 0.48 \\
                $L_2$ & Direct migration background (yes=1) &0.21 & 0.20 & 0.22 \\
                \hline
                $Y$&\textbf{Outcome:} Log gross hourly wage & 2.80 (2.45; 3.14)& 2.67 (2.34; 3.02)& 2.93 (2.60; 3.29)\\
                \hline
            \end{tabular}
               \end{adjustbox}
            \caption{Assumed mediators of the gender pay gap, covariates, and the outcome, with empirical means for both binary and numerical variables, and interquartile ranges (IQR) for numerical variables, presented for the total sample, women, and men.}
            \label{8Mediators}

\end{table}

The study population comprised 11,924 individuals after the exclusion of individuals without earnings or less than one euro, those under the age of 18 or above the retirement age of 67, pensioners, persons in education, and individuals with missing values.  Women (6,182, 51.8\%) were treated as the exposure group ($A=1$), while men (5,742) were treated as the control group ($A=0$). 
Women earned an average of 16.96 euros per hour, about5 euros less than men (21.94 euros). The marginal gender pay gap in this sample, defined as the percentage by which women earned less than men on average, was 22.7\%. This number is in line with the findings of the German Federal Statistical Office, which reported a gender pay gap of 22\% in Western Germany in 2017 \citep{Destatis_GenderPayGap_2022}.  The marginal gender pay gap in log gross hourly wages in the sample was 8.9\%.

Two strategies were considered with respect to the target group for mediator interventions. The first targeted women (Aims 1a-b), while the second targeted both women and men (Aims 2a-b). Since the education gap between women and men has narrowed significantly in recent decades \citep{Blau_Kahn_2017}, the mediator interventions in Aims 1a-b focused on aligning the other mediators, namely job characteristics and work experience. In order to reflect the relevance of age ($C$) and education ($M_1, M_2$) to career paths, the intervention distributions were conditioned on these factors. They were not conditioned on having a child ($L_1$) and a migration background ($L_2$) to avoid restricting women to the intervention distributions prevalent in the strata defined by these covariates among men. Specifically, the research questions are:

\begin{enumerate}
\item[Aim 1a] \textit{What reduction in the marginal gender pay gap in log gross hourly wages could be achieved if women’s job characteristics and work experience $(\boldsymbol{I}=M_3, \ldots, M_{10})$ were aligned with those of men within levels of age and educational background, while keeping the joint distribution of the other mediators in women constant as observed, given $C$ and $\boldsymbol{L}$?}
\item[Aim 1b] \textit{Which intervention would yield the largest reduction when setting the distribution of a single mediator $M_j$ $(j=3,\ldots, 10)$ in women equal to that in men, given age and educational background, while keeping the joint distribution of the other mediators in women constant as observed, given $C$ and $\boldsymbol{L}$?}
\end{enumerate}

Aims 2a-b addressed the potential reductions in the gender pay gap if a policy targeting both women and men equalized the distributions of the hypothesized mediators. Specifically, the research questions are:
\begin{enumerate}
\item[Aim 2a] \textit{What reduction in the marginal gender pay gap in log gross hourly wages could be achieved by aligning the joint distribution of $M_1, \ldots, M_{10}$ for women and men to match that of the total sample population?}
\item[Aim 2b] \textit{Which intervention would yield the largest reduction when setting the distribution of a single mediator $M_k$ $(k=1, \ldots, 10)$ in both women and men equal to that in the total sample population, while keeping the joint distribution of the other mediators in women and men constant as observed, given $C$ and $\boldsymbol{L}$?}
\end{enumerate}

For an overview, Table \ref{Estimands_Application} lists the intervention distributions of the manipulated mediators, the estimands, and their corresponding interpretations for all study aims. 
\begin{table}[H]
\centering
\resizebox{\textwidth}{!}{
\begin{tabular}{|l|l|l|l|}
\hline
  \thead{Mediators in $\boldsymbol{I}$} & \makecell{\thead{Intervention distribution of $\boldsymbol{I}$}} & \thead{Causal Estimand} & \thead{Interpretation} \\
\hline

 \makecell{Aim 1a:\\ $M_3,\ldots,M_{10}$\\ \\ Aim 1b: \\ $M_j,$ \\$j=3,\ldots, 10$}& \small{\makecell{$P_{\boldsymbol{I}|A=0, C, M_1, M_2}(\boldsymbol{i}|A=0,C, M_1, M_2)$}}& \makecell{$E[Y-Y_{\widetilde{\boldsymbol{I}}\widetilde{\boldsymbol{R}}|A=1,C, \boldsymbol{L}}|A=1]$} & \makecell{Expected change in $Y$ in \\women following intervention \\ (observed - counterfactual)}\\
\hline
\makecell{Aim 2a:\\ $\boldsymbol{M}=M_1, \ldots, M_{10}$\\ \\Aim 2b: \\$M_k,$ \\$k=1,\ldots, 10$}& \makecell{$P_{\boldsymbol{I}}(\boldsymbol{i})$}& \makecell{A $=$ \\$E[Y-Y_{\widetilde{\boldsymbol{I}}\widetilde{\boldsymbol{R}}|A=1,C, \boldsymbol{L}}|A=1]$\\ \\ B $=$ \\$E[Y_{\widetilde{\boldsymbol{I}}\widetilde{\boldsymbol{R}}|A=0,C, \boldsymbol{L}}-Y|A=0]$\\ \\ C $=$ A+B \\ {}} & \makecell{A: Expected change in $Y$ in \\ women following intervention\\
(observed - counterfactual)\\
B: Expected change in $Y$ in \\ men following intervention\\
(counterfactual - observed)\\
C: Overall reduction in the \\ gender pay gap }\\
\hline

\end{tabular}}
\caption{Manipulated mediators, intervention distributions, estimands, and their interpretations for Aims 1a-b and 2a-b. Relative wage changes for women and men were calculated by dividing the difference between observed and counterfactual (post-intervention) wages by the observed wage. For all aims, percentage reductions in the gender pay gap were calculated by multiplying the effects by $100/(E[Y|A=1]-E[Y|A=0])$.}
\label{Estimands_Application}
\end{table}

The estimation of the expected counterfactual outcomes is based on the assumptions of positivity, consistency, and the absence of unmeasured mediator-outcome confounders. Notably, unmeasured common causes of the mediators (e.g., individual skills) should not bias the results. See Appendix, Section \ref{Contextualizing Aims}, for a summary of the required assumptions.

\subsection{MC g-computation for estimation}
\label{MC g-computation}
This analysis employed MC g-computation, following the method outlined by \citet{Daniel2017} for interventional (in)direct effects. The OB decomposition described in Section \ref{Section Oaxaca} and a parametric approach based on simple linear outcome models, as outlined by \citet{Jackson}, were deemed unsuitable for estimating the target estimands of Aims 1a-b and 2a-b. This is because wage models likely involve complex relationships, including various interaction terms and nonlinear effects. Furthermore, the OB decomposition does not address the targeted estimands. The intervention distributions for Aims 1a-b are conditional on covariates and, as a result, are not equivalent to marginal means. Aims 2a-b address interventions for both the exposure and control groups, rather than solely for the exposed. 

G-computation relies on correct model specification; however, it offers flexibility in modeling. Specifically, MC g-computation employs a simulation approach to generate counterfactual outcomes. This is achieved by randomly drawing values for each individual’s mediator and outcome variables, under the specified interventions, from fitted statistical models, thus generating counterfactual data sets. The algorithm for estimating $E[Y_{\widetilde{\boldsymbol{I}}\widetilde{\boldsymbol{R}}|A=a,C, \boldsymbol{L}}|A=a]$, $a\in \{0,1\}$, is outlined in Table \ref{Algo}. It adapts the estimator described by \citet{Daniel2017} in one main aspect: both the counterfactual outcome and the mediator intervention distributions are conditional on $A$, meaning that no interventions are applied to the exposure. For interventional effects involving exposure interventions, Table \ref{Algo_IIE} in the Appendix outlines the steps for estimating $E[Y_{a \boldsymbol{\widetilde{I}}_{a^*|\boldsymbol{C}} \boldsymbol{\widetilde{R}}_{a| \boldsymbol{C}}}]$, which is central to $\text{IIE}_{\boldsymbol{I}|\boldsymbol{C}}$ described in equation \ref{IIE}. 

In the concrete analysis, the joint distribution of mediators was specified using factorization based on the working order in Table \ref{8Mediators}. Logit models were fitted for binary variables and linear models for continuous variables. All possible two-way interaction terms among the regressors have been considered in the model-building process for the outcome model and the joint mediator distribution. The models were chosen depending on Akaike's model selection criterion. \citet{Daniel2017} suggested drawing several million times from the fitted models to ensure that the results are free of Monte Carlo error to the number of decimal places given. This analysis achieved stable results at 300 draws per unit. The algorithm was therefore implemented with 300 Monte Carlo runs (i.e., $Z=300$ in Table \ref{Algo}). 95\% confidence intervals were obtained using the nonparametric bootstrap, with 1000 bootstrap samples (i.e., $B=1000$ in Table \ref{Algo}).

A side note on the log-transformed outcomes: The g-computation method generates counterfactual datasets with outcomes for each individual, which can be converted back to the original scale \citep[as done, e.g., in][]{Schomaker_2017_GComp}.  Nevertheless, as the log transformation of wages is standard in econometric literature and gender pay gap research \citep{Oaxaca1973, Blinder1973, Wooldridge_2010_Econometrics, Mischler2021Verdienstunterschiede, schmitt_stall_2022}, this article presents results based on log hourly wages in order to enhance comparability.

\begin{table}
{
\renewcommand{\arraystretch}{1.1}
\begin{tabular}{p{0.095\textwidth}|p{0.875\textwidth}}
\hline
\multicolumn{2}{l}{\textbf{Algorithm to estimate $E[Y_{\widetilde{\boldsymbol{I}}\widetilde{\boldsymbol{R}}|A=a,C, \boldsymbol{L}}|A=a]$, $a\in \{0,1\}$ }}\\
\hline
     Step 1 & Specify and fit models for:\\
        & 1) the distribution of $\boldsymbol{I}$; for Aims 1a-b: conditional on $A=0$, $C$, $M_1$ and $M_2$, for Aim 2 a-b: the distribution of $\boldsymbol{I}$ in the overall sample (Model 1) \\
         & 2) the distribution of $\boldsymbol{R}$ conditional on $A=a$,  $C$ and $\boldsymbol{L}$ (Model 2),\\
         & 3) the outcome $Y$ in conditional on $A=a$,  $C$, $\boldsymbol{L}$ and $\boldsymbol{M}$ (Model 3). \\
        & For each individual with $A=a$:\\
     Step 2 & Set the distribution of $C$ and $\boldsymbol{L}$ to the respective empirical distributions in $A=a$. \\
    Step 3 & Set the variables in $\boldsymbol{I}$ to a random draw from Model 1 fitted in Step 1.\\
         Step 4 & Set the variables in $\boldsymbol{R}$ to a random draw from Model 2 fitted in Step 1.\\
    Step 5 & Draw the outcome from Model 3 fitted in Step 1 based on the updated data generated through steps 2-4. \\
     Step 6 & Estimate $E[Y_{\widetilde{\boldsymbol{I}}\widetilde{\boldsymbol{R}}|A=a,C, \boldsymbol{L}}|A=a]$ by calculating the mean of the simulated outcomes in step 5. \\
      Step 7 & Repeat steps 3-6 Z times with different seeds and calculate the mean of the Z estimates. \\
    Step 8 & To obtain the nonparametric bootstrap standard error and percentile confidence intervals, repeat steps 1-7 $B$ times on the bootstrapped data. \\
     \hline
\end{tabular}
}
\caption{Proposed algorithm to estimate $E[Y_{\widetilde{\boldsymbol{I}}\widetilde{\boldsymbol{R}}|A=a,C, \boldsymbol{L}}|A=a]$,  $a\in \{0,1\}$.}
\label{Algo}
\end{table}

\subsection{Results}
\label{Applied analysis: Results}

\subsubsection{Aims 1a-b}
\label{Applied analysis: Results Aim1}

\textbf{Aim 1a:} The marginal gender pay gap in log gross hourly wages (8.9\%, 95\% confidence interval (CI): 8.2; 9.4) was estimated to be reduced by 71.2\% (CI: 63.4; 80.2), when setting the joint distribution of job characteristics and work experience ($M_3, \ldots, M_{10}$) given age and educational background among women equal to that among men, while holding the joint distribution of the other intermediates constant. This is attributed to an estimated increase of  6.9\% (CI: 6.2; 7.7) in the average log wages of women. As a result, a gender pay gap of 2.5\% (CI: 1.7; 3.4) would remain.\\
\textbf{Aim 1b:} When prioritizing a single-mediator intervention, the marginal gender pay gap in log gross hourly wages would be most reduced, by 26.5\% (CI: 20.9; 32.6), if women had the same distribution of work experience as men within strata of age and educational background. This is attributed to an estimated increase of 2.6\% (CI: 2.0; 3.1) in the average log wages of women. As a result, a gender pay gap in log gross hourly wages of 6.5\% (CI: 5.7; 7.2) would remain.    

\subsubsection{Aims 2a-b}
\label{Applied analysis: Results Aim2}
\textbf{Aim 2a:} The marginal gender pay gap in log gross hourly was estimated to be reduced by 85.6\% (CI: 69.4; 104.0), if the joint distribution of all mediators $M_1, \ldots, M_{10}$ in women and in men were set equal to that in the overall sample. 59.4\% (CI: 49.2; 68.6) of this reduction 
is due to a decrease in log wages in men (4.5\% decrease, CI: 3.2; 5.8), while the remaining 40.6\% (CI: 31.4; 50.8) is due to an increase in log  wages in women (3.4\% increase, CI: 2.5; 4.3). As a result, a gender pay gap of  1.3\% (CI: -0.4; 2.8) would remain. \\
\textbf{Aim 2b:} When prioritizing a single-mediator intervention, the marginal gender pay gap in log gross hourly wages would be most reduced, by 29.8\% (CI: 20.2; 40), if the distribution of full-time employment in women and in men matched that in the overall sample. 90\% (CI: 80.7; 98.2) of this reduction is due to a decline in log wages in men (2.4\% decrease, CI: 1.6; 3.2), while the remaining 10\% (CI: 1.8; 19.3) is due to a slight increase in women's log wages  (0.3\% increase, CI: 0.1; 0.5). As a result, the remaining gender pay gap would be 6.4\% (CI: 5.3; 7.4). The second largest reduction in the pay gap (21.5\%, CI: 9.2; 35.3) could be achieved by aligning the distribution of work experience with that of the overall sample, mainly due to an increase in women's wages. In contrast, aligning educational backgrounds, which would lead to wage losses for women and gains for men, would widen the gap (see Figure \ref{Grafik1_Results_2b} and Table \ref{Aim2b_Results} in the Appendix for detailed results).
 
 \begin{figure}[H]
    \centering
    \includegraphics[width=1\textwidth]{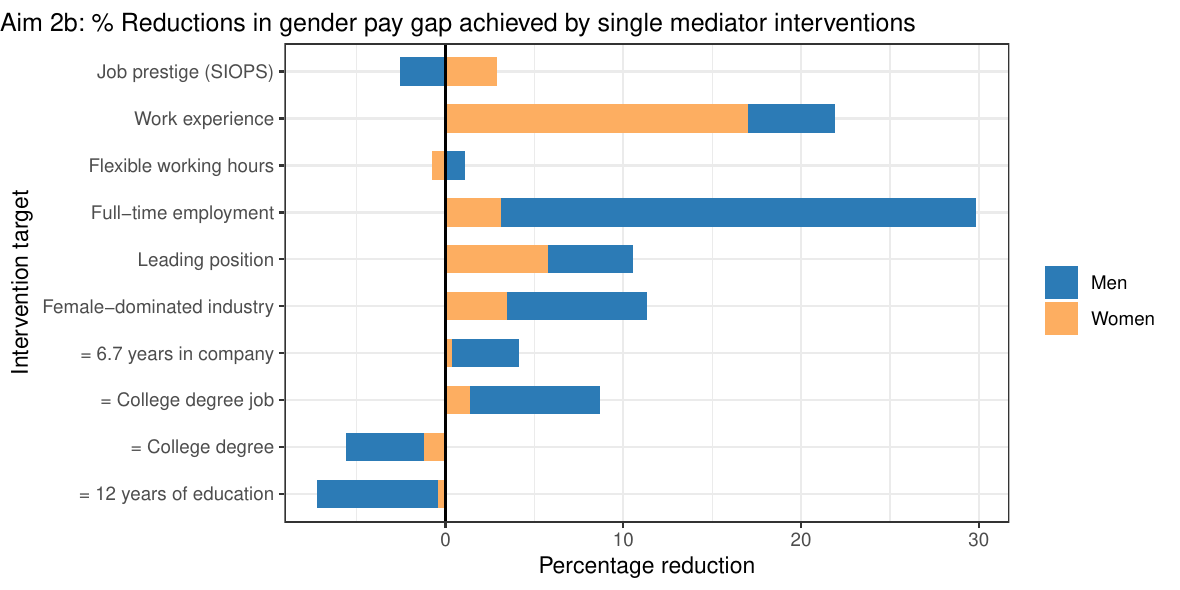}
    \caption{Percentage reductions in the marginal gender pay gap in log gross hourly wages achieved by setting the distribution of a single mediator among women and among men equal to that observed in the overall sample (Aim 2b), with portions attributable to changes in log wages for women and men. Positive values indicate a reduction in the pay gap, while negative values indicate an increase. For men, positive percentages reflect decreased average log wages, while for women they indicate increases; the opposite is true for negative values.}
    \label{Grafik1_Results_2b}
\end{figure}

\section{Discussion}
\label{Section Discussion}
This article has outlined three distinct approaches to the mediating mechanisms of social disparities, each illustrated by representative target estimands. Approach 1 focuses on causal mediation to investigate the mechanisms underlying a social disparity.  Approaches 2 and 3 assess the effects of hypothetical manipulations of mediators: Approach 2 on the exposure-induced disparity, and Approach 3 on the actual observed disparity. As such, both approaches are policy-relevant for identifying potential intervention targets to reduce social disparities \citep{Jackson, Moreno_2021_emulate_target}. Approaches 1 and 2 entail the construction of causal effects of the social exposure, which imply hypothetical interventions on the exposure. However, when the exposure is a complex or non-manipulable construct, such as gender or race, defining meaningful interventions on the exposure becomes challenging. In such instances, Approach 3, which does not entail the manipulation of the exposure, may offer more clearly defined target quantities. With regard to other exposure variables, such as membership of an educational group or social organization, causal effects of the exposure may be less challenging to conceptualize, thereby rendering Approaches 1 or 2 viable options. Regarding the strength of the required conditions, Approach 1 is the most demanding. In the context of mediation analysis, the key measures for causal mechanisms are typically natural (in)direct effects \citep{nguyen_clarifying_2021, Pearl2001}. However, these measures rely on strong identification assumptions, including cross-world assumptions, which are often violated in multiple-mediator settings. Alternative indirect effect measures, including interventional indirect effects, may fail to capture the pathways under scrutiny \citep[][]{Miles_2023}. A novel approach to mediation has recently been proposed by \citet{Diaz_nonagency_2023}, showing promise for future research on the mechanisms of social disparities. This approach enables the identification and estimation of the strength of causal mediating mechanisms, even in the presence of exposure-induced mediator-outcome confounders. The proposed path-specific effects are based on non-agency interventions, a useful concept for non-manipulable exposures. Moreover, these effects are zero when no mediating mechanism exists, thus satisfying the path-specific sharp null criteria. 

The paper has placed a variant of the prominent twofold OB decomposition for linear outcome models within Approaches 1-3, concluding that causal implementations of the OB method are limited to estimating specific estimands that require strong identification assumptions. Moreover, when the model for the outcome (e.g., wage) involves a complex functional form, including nonlinearities or higher-order interactions, other estimation methods that allow for flexible modeling, such as MC g-computation, are considered more suitable.

MC g-computation has been employed in Approach 3 to analyze the gender pay gap in Western Germany, using data from the 2017 SOEP. The focus has been on hypothetical interventions targeting potential mediators of the gender pay gap. Specifically, the analysis has examined potential reductions in the observed gap by altering mediator distributions. When interpreting the results, it is important to note that the findings relate to changes in the gender pay gap in log gross hourly wages. Due to the nonlinear transformation, these results do not directly translate to changes in the gap in actual hourly wages. 

The results for the first study aim suggest that approximately 71\% of the gender pay gap could be reduced if women had the same age- and education-specific distribution of job characteristics and work experience as men. In particular, a substantial portion of the gap could be reduced if women had the same distribution of work experience as men. This may reflect wage penalties women face due to career interruptions, such as those related to parenthood and family care \citep{Kleven_2019_ChildPenalties, Ehrlich_2020_CareWork_Employment}. 
These findings are consistent with those of previous studies (see \citep{Blau_Kahn_2017, kunze2018gender} for an overview of existing literature). However, there is considerable variation in the literature on the gender pay gap, particularly with regard to the set of variables, data sources, time periods, regions, and analytical approaches. This variability complicates direct comparisons. For example, a decomposition analysis conducted by the German Federal Statistical Office, based on the 2018 Structure of Earnings Survey, attributed approximately 71\% of the gender pay gap in log gross hourly wages to gender differences in the marginal means of wage-determining factors \citep{Mischler2021Verdienstunterschiede}. This analysis used different data and variables than this article, covered all of Germany, employed the OB method, and did not adopt a clear causal approach, despite being labeled a ``cause analysis''.

The results of the second study aim differ from those of the first, highlighting the importance of how estimands are defined. Specifically, the findings suggest that the gender pay gap could be reduced by approximately 86\% if the joint distribution of all considered mediators among women and men were to align with that of the total sample population.
The single-mediator analysis indicates that the gender pay gap could be substantially smaller if the proportion of men and women in full-time employment matched that of the full sample, while the joint distributions of other wage-relevant factors remained unchanged. Although women would experience a larger relative increase in full-time employment compared to the decrease for men as a result of this alignment, the relative wage gain for women would be modest, while the wage loss for men would be comparatively large. This could be due to a part-time wage penalty for men, as found in previous studies \citep{Hirsch_US_2005_Parttime_Job_characeristics, Russo_Parttime_Penalty_2005, O'Dorchai_Parttime_Men_2007, wolf2014_ParttimePenalty, Nightingale_ParttimeMen_2019}. One possible explanation for a relatively high part-time penalty for men is that part-time work is less common among men than women, making it more noticeable. This may lead to lower wages, as employers could perceive men's part-time work as a signal of reduced work commitment \citep{Russo_Parttime_Penalty_2005}.

Further research is needed to gain more nuanced insights into the gender-specific effects of mediator interventions on wages. One potential avenue of research would be to examine interventions in men and women with comparable covariates, such as equalizing full-time employment within the same occupations, while accounting for industry-specific wage structures (e.g., mandatory wage scales or the role of wage negotiations \citep[][]{ANTONCZYK2010_collectivebargaining,BabcockLaschever+2003, Dittrich_WAgeNegotiations_2014, exley2020knowing}). Defining a reasonable and comprehensive set of covariates on which the stochastic assignment of mediator values is conditioned is also crucial to achieve realistic interventions at the individual level. Without a meaningful selection of covariates, there is a risk of assigning implausible mediator values to individuals. 

Approach 3 has the fewest prerequisites among the approaches discussed in Section \ref{Section: Interventional effects general}. However, the identification of relevant counterfactual outcomes still relies on several key conditions, in particular: the absence of unmeasured confounders in mediator-outcome relationships, and assumptions of positivity and consistency. In the context of the gender pay gap, unmeasured confounders may include wage determinants that are not included in the analysis. Positivity violations occur when women or men have a zero probability of receiving a mediator value in the support of the intervention distribution, while consistency may be compromised if mediator interventions are not clearly defined. Additionally, real shifts in mediator distributions (e.g., more men working part-time) could affect working conditions for both genders in complex ways that are not captured by the causal model. This complicates the estimation of the real-world impact of mediator changes on the gender pay gap.
Beyond the empirical example, the paper concludes that the interventional effects framework, as discussed in various contexts including mediation, is a useful tool in social inequality research. In particular, the identification of targets for reducing social inequalities is a useful application of the framework.

\subsection*{Acknowledgements}
The author is grateful for the reviewers' valuable comments, which improved the manuscript, and to Josef Br\"uderl and Michael Schomaker for their feedback on the early draft of the manuscript.
\subsection*{Funding information}
The author states no funding involved.
\subsection*{Conflict of interest}
The author states no conflict of interest.
\subsection*{Author contribution}
The author confirms the sole responsibility for the conception of the study, presented results and manuscript preparation.
\subsection*{Data availability statement}
The data that support the findings of this study are available from the German Institute for Economic Research (DIW Berlin) but restrictions
apply to the availability of these data, which were used under license for the current study, and so are
not publicly available. Data are however available from the author upon reasonable request and with
permission of the DIW Berlin.
\subsection*{Ethical approval}
The author used secondary data from the German Socio-Economic Panel (SOEP). The SOEP places the highest priority on protecting the confidentiality of respondents’ data by
ensuring strict adherence to European and German data protection regulations, consistent with the principles of the Helsinki Declaration. 

\newpage
\addcontentsline{toc}{section}{References}
\bibliographystyle{unsrtnat}
{\footnotesize

}
\newpage

\appendix
\begin{center}
    \LARGE{\textbf{Appendix}}
\end{center}

\section{Structural assumptions including unmeasured common causes of the exposure, intermediates and the outcome}

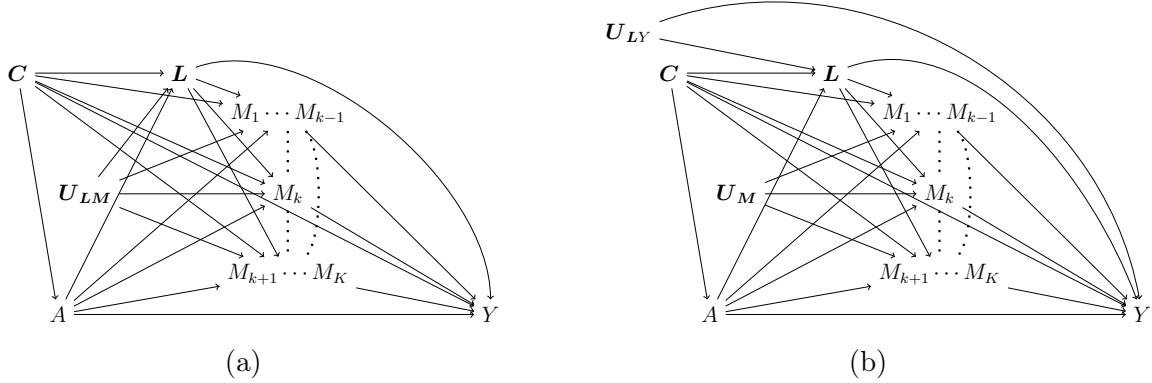
\begin{figure}[H]
\subfloat[]{
 \centering
 \scalebox{0.7}{
\begin{tikzpicture}[x=10in,y=6in]
\node (M1) at (0.15,-0.050) {$M_1 \cdots M_{k-1}$};
\node (A) at (-0.02,-0.300) {$A$};
\node (U_LM) at (0.00,-0.150) {$\boldsymbol{U}_{\boldsymbol{L}\boldsymbol{M}}$};
\node (M7) at (0.15,-0.150) {$M_k$};
\node (M10) at (0.15,-0.250) {$M_{k+1} \cdots M_{K}$};
\node (Y) at (0.3,-0.300) {$Y$};
\node (C) at (-.05,-0.0) {$\boldsymbol{C}$};
\node (L) at (0.07,0.0) {$\boldsymbol{L}$};

\draw [->] (U_LM) edge (M1);
\draw [->] (U_LM) edge (M7);
\draw [->] (U_LM) edge (M10);
\draw [->] (U_LM) edge (L);
\draw [->] (A) edge (M1);
\draw [->] (A) edge (M7);
\draw [->] (A) edge (M10);
\draw [->] (M1) edge (Y);
\draw [->] (M7) edge (Y);
\draw [->] (M10) edge (Y);
\draw [->] (C) edge (M1);
\draw [->] (C) edge (M7);
\draw [->] (C) edge (M10);
\draw [->] (C) edge (Y);
\draw [->] (C) edge (A);
\draw [->] (A) edge (Y);
\draw [->] (A) edge (L);
\draw [->] (L) edge (M1);
\draw [->] (L) edge (M7);
\draw [->] (L) edge (M10);
\draw [->] (L) to [out=20,in=90, looseness=0.8] (Y);

\draw [->] (C) edge (L);

\draw [loosely dotted, very thick]  (M1) edge (M7);
\draw [loosely dotted, very thick] (M7) edge (M10);
\draw [loosely dotted, very thick] (M1) to [out=-45,in=45, looseness=0.5] (M10);
\end{tikzpicture}
}

}
\hspace*{0.6cm}
  \subfloat[]{
 \centering
 \scalebox{0.7}{

\begin{tikzpicture}[x=10in,y=6in]
\node (M1) at (0.15,-0.050) {$M_1 \cdots M_{k-1}$};
\node (A) at (-0.02,-0.300) {$A$};
\node (U_M) at (0.00,-0.150) {$\boldsymbol{U}_{\boldsymbol{M}}$};
\node (U_L) at (-0.08,0.05) {$\boldsymbol{U}_{\boldsymbol{L}Y}$};
\node (M7) at (0.15,-0.150) {$M_k$};
\node (M10) at (0.15,-0.250) {$M_{k+1} \cdots M_{K}$};
\node (Y) at (0.3,-0.300) {$Y$};
\node (C) at (-.05,-0.0) {$\boldsymbol{C}$};
\node (L) at (0.07,0.0) {$\boldsymbol{L}$};

\draw [->] (U_M) edge (M1);
\draw [->] (U_M) edge (M7);
\draw [->] (U_M) edge (M10);
\draw [->] (A) edge (M1);
\draw [->] (A) edge (M7);
\draw [->] (A) edge (M10);
\draw [->] (M1) edge (Y);
\draw [->] (M7) edge (Y);
\draw [->] (M10) edge (Y);
\draw [->] (C) edge (M1);
\draw [->] (C) edge (M7);
\draw [->] (C) edge (M10);
\draw [->] (C) edge (Y);
\draw [->] (C) edge (A);
\draw [->] (A) edge (Y);
\draw [->] (A) edge (L);
\draw [->] (L) edge (M1);
\draw [->] (L) edge (M7);
\draw [->] (L) edge (M10);
\draw [->] (L) to [out=20,in=110, looseness=0.9] (Y);
\draw [->] (U_L) to (L);
\draw [->] (U_L) to [out=20,in=100, looseness=1] (Y);
\draw [->] (C) edge (L);

\draw [loosely dotted, very thick]  (M1) edge (M7);
\draw [loosely dotted, very thick] (M7) edge (M10);
\draw [loosely dotted, very thick] (M1) to [out=-45,in=45, looseness=0.5] (M10);

\end{tikzpicture}
}
}

        \caption{Structural assumptions encompassing unmeasured common causes  of the intermediate factors in $\boldsymbol{L}$ and $\boldsymbol{M}$ ($\boldsymbol{U_{LM}}$) (a), or unmeasured common causes of $\boldsymbol{L}$ and $Y$ ($\boldsymbol{U}_{\boldsymbol{L}Y}$), and of the mediators in $\boldsymbol{M}$ ($\boldsymbol{U}_{\boldsymbol{M}}$) (b), under which assumptions \ref{A1}-\ref{A3} (main text) hold.}
        \label{Structural assumptions unmeasured common causes of intermediates}
\end{figure}

\tikzset{
  big arrow/.style={
    decoration={markings,mark=at position 1 with {\arrow[scale=4,#1]{>}}},
    postaction={decorate},
    shorten >=0.4pt},
  big arrow/.default=blue}

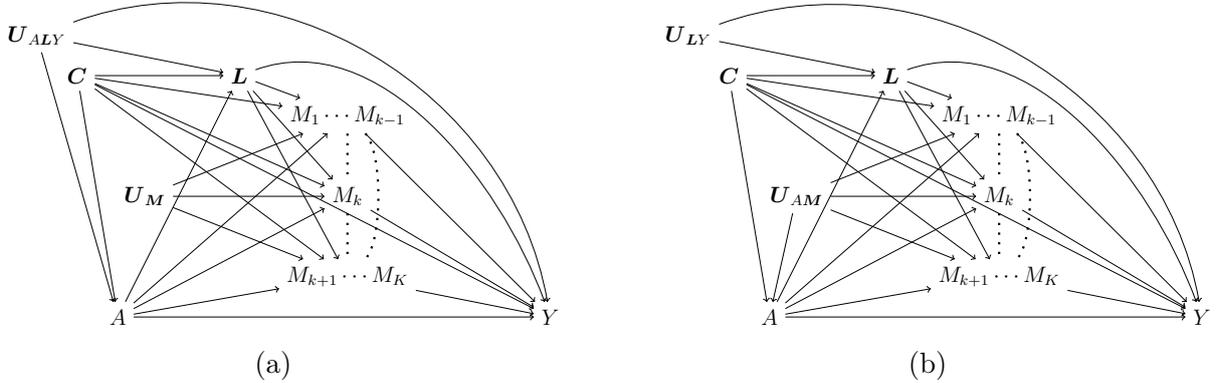
\begin{figure}[H]
\subfloat[]{
 \centering
 \scalebox{0.7}{
\begin{tikzpicture}[x=10in,y=6in]
\node (M1) at (0.15,-0.050) {$M_1 \cdots M_{k-1}$};
\node (A) at (-0.02,-0.300) {$A$};
\node (U_M) at (0.00,-0.150) {$\boldsymbol{U}_{\boldsymbol{M}}$};
\node (U_L) at (-0.08,0.05) {$\boldsymbol{U}_{A\boldsymbol{L}Y}$};
\node (M7) at (0.15,-0.150) {$M_k$};
\node (M10) at (0.15,-0.250) {$M_{k+1} \cdots M_{K}$};
\node (Y) at (0.3,-0.300) {$Y$};
\node (C) at (-.05,-0.0) {$\boldsymbol{C}$};
\node (L) at (0.07,0.0) {$\boldsymbol{L}$};

\draw [->] (U_M) edge (M1);
\draw [->] (U_M) edge (M7);
\draw [->] (U_M) edge (M10);
\draw [->] (A) edge (M1);
\draw [->] (A) edge (M7);
\draw [->] (A) edge (M10);
\draw [->] (M1) edge (Y);
\draw [->] (M7) edge (Y);
\draw [->] (M10) edge (Y);
\draw [->] (C) edge (M1);
\draw [->] (C) edge (M7);
\draw [->] (C) edge (M10);
\draw [->] (C) edge (Y);
\draw [->] (C) edge (A);
\draw [->] (A) edge (Y);
\draw [->] (A) edge (L);
\draw [->] (L) edge (M1);
\draw [->] (L) edge (M7);
\draw [->] (L) edge (M10);
\draw [->] (L) to [out=20,in=110, looseness=0.9] (Y);
\draw [->] (U_L) to (L);
\draw [->] (U_L) to [out=20,in=100, looseness=1] (Y);
\draw [->] (U_L) to (A);
\draw [->] (C) edge (L);

\draw [loosely dotted, very thick]  (M1) edge (M7);
\draw [loosely dotted, very thick] (M7) edge (M10);
\draw [loosely dotted, very thick] (M1) to [out=-45,in=45, looseness=0.5] (M10);

\end{tikzpicture}
}

}
\hspace*{0.6cm}
  \subfloat[]{
 \centering
 \scalebox{0.7}{

\begin{tikzpicture}[x=10in,y=6in]
\node (M1) at (0.15,-0.050) {$M_1 \cdots M_{k-1}$};
\node (A) at (-0.02,-0.300) {$A$};
\node (U_M) at (0.00,-0.150) {$\boldsymbol{U}_{A\boldsymbol{M}}$};
\node (U_L) at (-0.08,0.05) {$\boldsymbol{U}_{\boldsymbol{L}Y}$};
\node (M7) at (0.15,-0.150) {$M_k$};
\node (M10) at (0.15,-0.250) {$M_{k+1} \cdots M_{K}$};
\node (Y) at (0.3,-0.300) {$Y$};
\node (C) at (-.05,-0.0) {$\boldsymbol{C}$};
\node (L) at (0.07,0.0) {$\boldsymbol{L}$};

\draw [->] (U_M) edge (M1);
\draw [->] (U_M) edge (M7);
\draw [->] (U_M) edge (M10);
\draw [->] (U_M) edge (A);
\draw [->] (A) edge (M1);
\draw [->] (A) edge (M7);
\draw [->] (A) edge (M10);
\draw [->] (M1) edge (Y);
\draw [->] (M7) edge (Y);
\draw [->] (M10) edge (Y);
\draw [->] (C) edge (M1);
\draw [->] (C) edge (M7);
\draw [->] (C) edge (M10);
\draw [->] (C) edge (Y);
\draw [->] (C) edge (A);
\draw [->] (A) edge (Y);
\draw [->] (A) edge (L);
\draw [->] (L) edge (M1);
\draw [->] (L) edge (M7);
\draw [->] (L) edge (M10);
\draw [->] (L) to [out=20,in=110, looseness=0.9] (Y);
\draw [->] (U_L) to (L);
\draw [->] (U_L) to [out=20,in=100, looseness=1] (Y);
\draw [->] (C) edge (L);

\draw [loosely dotted, very thick]  (M1) edge (M7);
\draw [loosely dotted, very thick] (M7) edge (M10);
\draw [loosely dotted, very thick] (M1) to [out=-45,in=45, looseness=0.5] (M10);

\end{tikzpicture}
}
}

        \caption{Structural assumptions encompassing unmeasured common causes of $A$, $\boldsymbol{L}$ and $Y$ ($\boldsymbol{U}_{A\boldsymbol{L}Y}$), and of the mediators in $\boldsymbol{M}$ ($\boldsymbol{U}_{\boldsymbol{M}}$) (a), or of $A$ and $\boldsymbol{M}$ ($\boldsymbol{U}_{A\boldsymbol{M}}$), and $\boldsymbol{L}$ and $Y$ ($\boldsymbol{U}_{\boldsymbol{L}Y}$) (b), under which $\text{IE}_{\boldsymbol{I}|\boldsymbol{C}(obs)}$ (equation \ref{IEobsI}) is identified.}
        \label{Structural assumptions to illustrate Approaches3}
\end{figure}
\section{Alternative indirect effect measures in the presence of exposure-induced confounders}
\label{AlternativeEffectmeasures}
When operating under the DAG in Figure \ref{Structural assumptions to illustrate Approaches 1-3} of the main text,
  one potential approach is to consider all intermediate factors of the causal model, i.e., $\boldsymbol{L}$ and $\boldsymbol{M}$, in a single mediator set to evaluate the natural indirect effect through $\boldsymbol{L}$ and $\boldsymbol{M}$, as proposed by \citet{Vansteelandt_MultipleMediators2014} and \citet{VanderWeeleVansteeland_Robins_RIA2014}. This effect would be equivalent to the interventional indirect effect through $\boldsymbol{L}$ and $\boldsymbol{M}$, since all exposure-induced confounders are included in the mediator set \citep[][]{Miles_2023}. However, considering all intermediates jointly is not a suitable approach for investigating indirect pathways through specific mediators. Consequently, this approach does not address \ref{Question 1} in the main text. An alternative option could be to consider the path-specific indirect effect $A \rightarrow \boldsymbol{M} \rightarrow Y$, which does not involve the paths from $A$ to $\boldsymbol{M}$ via exposure-induced confounders $\boldsymbol{L}$ \citep{Avin2005}. It can be defined as an interventional path-specific indirect effect where the intervention distribution of $\boldsymbol{M}$ is conditional on $\boldsymbol{C}$ and $\boldsymbol{L}_{0}=l$, as suggested in \citet{ NguyenSchmidOgburnStuart_2022, Zheng_Laan_Survival_2017, Miles_2023}. This effect offers a mediational interpretation, but is not an interventional analogue to the natural indirect effect through all mediators in $\boldsymbol{M}$, which includes $A \rightarrow \boldsymbol{M} \rightarrow Y$ and $A \rightarrow \boldsymbol{L} \rightarrow \boldsymbol{M} \rightarrow Y$  \citep{Miles_2023}. Path-specific indirect effects through a proper mediator subset  $\boldsymbol{I}$, which do not involve the pathways from $A$ to $\boldsymbol{I}$ via intermediates prior to $\boldsymbol{I}$, necessitate specifying the causal order among $\boldsymbol{I}$ and the other mediators in $\boldsymbol{R}$  \citep{Daniel2017, LinVanderWeele_PathSpecific_2017}. Thus, they are not alternatives to the estimand targeted by \ref{Question 1} when the causal order among the mediators is unspecified.

\section{Identification results}
\label{Identification}
\citet{NguyenSchmidOgburnStuart_2022} provide identification proofs for counterfactual outcomes when the mediator intervention distribution is a counterfactual distribution, including distributions that are either marginal or conditional on covariates $\boldsymbol{C}$, or on  $\boldsymbol{C}$ and $\boldsymbol{L}_a$. The results can be applied to the identification of $\text{IIE}_{\boldsymbol{I}|\boldsymbol{C}} =  E[Y_{1 \boldsymbol{\widetilde{I}}_{1|\boldsymbol{C}}\boldsymbol{\widetilde{R}}_{1|\boldsymbol{C}}}-Y_{1 \boldsymbol{\widetilde{I}}_{0|\boldsymbol{C}} \boldsymbol{\widetilde{R}}_{1| \boldsymbol{C}}}]$ (equation \ref{IIE}), $\text{IE}_{\boldsymbol{I}|\boldsymbol{C}} =  E[Y_{1 \boldsymbol{M}_1}-Y_{1 \boldsymbol{\widetilde{I}}_{0|\boldsymbol{C}} \boldsymbol{\widetilde{R}}_{1|\boldsymbol{C}, \boldsymbol{L}_1}}]$ (equation \ref{IE_I}) of the main text. Thus, only the derivations of the intervention distributions $P_{\boldsymbol{I}_{a|\boldsymbol{C}}}(\boldsymbol{i}|\boldsymbol{C})$ and  $P_{\boldsymbol{R}_{1|\boldsymbol{C}, \boldsymbol{L}_1}}(\boldsymbol{r}|\boldsymbol{C}, \boldsymbol{L}_1)$ are provided here. These derivations closely follow those in \citet[][]{NguyenSchmidOgburnStuart_2022}. As a notation reminder, $\boldsymbol{I}$ and $\boldsymbol{R}$ are subsets of $\boldsymbol{M}$, i.e., $\boldsymbol{M}=\boldsymbol{I}\cup\boldsymbol{R}=(M_1,\ldots,M_K)$.
\vspace{0.5cm}

For any possible mediator subset $\boldsymbol{I} \subseteq \boldsymbol{M}$, we have for $P_{\boldsymbol{I}_{a|\boldsymbol{C}}}(\boldsymbol{i}|\boldsymbol{C}) = P(\boldsymbol{I}_a = \boldsymbol{i} | \boldsymbol{C})$, $a\in \{0,1\}$:
\begin{align}
    P(\boldsymbol{I}_a = \boldsymbol{i} | \boldsymbol{C}) &= P(\boldsymbol{I}_a = \boldsymbol{i}  | \boldsymbol{C}, A = a) &\quad {\text{conditional independence} \; (M_{1a}, \ldots M_{Ka}) \independent A | \boldsymbol{C}}, \\ 
    &= P(\boldsymbol{I} = \boldsymbol{i}  | \boldsymbol{C}, A = a) &\quad {\text{consistency \& positivity}}.
\end{align}

By the weak union rule of conditional independence, the assumption $(\boldsymbol{L}_a, M_{1a}, \ldots, M_{Ka}) \independent A | \boldsymbol{C}$ implies that $(M_{1a}, \ldots, M_{Ka}) \independent A | \boldsymbol{C}, \boldsymbol{L}_a$ \citep{NguyenSchmidOgburnStuart_2022}. Thus, for $P_{\boldsymbol{R}_{1|\boldsymbol{C}, \boldsymbol{L}_1}}(\boldsymbol{r}|\boldsymbol{C}, \boldsymbol{L}_1)=P(\boldsymbol{R}_1 = \boldsymbol{r} | \boldsymbol{C}, \boldsymbol{L}_1)$, we have:
\begin{align}
    P(\boldsymbol{R}_1 = \boldsymbol{r} | \boldsymbol{C}, \boldsymbol{L}_1) &= P(\boldsymbol{R}_1 = \boldsymbol{r}  | \boldsymbol{C}, \boldsymbol{L}_1, A = 1) &\quad {\text{conditional independence}} \\
    & & {(M_{1a}, \ldots, M_{Ka}) \independent A | \boldsymbol{C}, \boldsymbol{L}_a} \nonumber \\
    &= P(\boldsymbol{R} = \boldsymbol{r}  | \boldsymbol{C}, \boldsymbol{L}=\boldsymbol{l}, A = 1) &\quad {\text{consistency \& positivity}}.
\end{align}

\vspace{0.5cm}
Identification results for $E[Y_{1\widetilde{\boldsymbol{I}}|A=0, \boldsymbol{C} {\widetilde{\boldsymbol{R}}|A=1,\boldsymbol{C}, \boldsymbol{L}}}|A=1]$ in  $\text{IE}_{\boldsymbol{I}|\boldsymbol{C}(obs)}$ (equation \ref{IEobsI}) build on the proofs
in \citet[][]{NguyenSchmidOgburnStuart_2022, Jackson}. Here, the counterfactual outcome is defined to be conditional on $A=1$ and the mediator intervention distributions are functions of the observed data distribution. 
{\footnotesize{
\begin{align}
 &E[Y_{\widetilde{\boldsymbol{I}}|A=0, \boldsymbol{C} \,{\widetilde{\boldsymbol{R}}|A=1,\boldsymbol{C}, \boldsymbol{L}}}|A=1] =  E \left( E \left\{ E[Y_{\widetilde{\boldsymbol{I}}|A=0, \boldsymbol{C} \,{\widetilde{\boldsymbol{R}}|A=1,\boldsymbol{C}, \boldsymbol{L}}}|A=1, \boldsymbol{C}, \boldsymbol{L}]|A=1, \boldsymbol{C}| A=1 \right\}\right)     \\
 &{\text{iterated expectation}} \nonumber \\ 
&=\sum_{\boldsymbol{c}} \sum_{\boldsymbol{l}} \sum_{\boldsymbol{m}}  E[Y_{\boldsymbol{i}\boldsymbol{r}}| A=1, \boldsymbol{c},\boldsymbol{l}] P(\boldsymbol{i}|A=0, \boldsymbol{c})P(\boldsymbol{r}|A=1, \boldsymbol{c},   \boldsymbol{l}) P(\boldsymbol{l}|A=1, \boldsymbol{c})  P(\boldsymbol{c}|A=1) \\
 &{\text{transition to summation}}\footnotemark[1] \nonumber \\
&=\sum_{\boldsymbol{c}} \sum_{\boldsymbol{l}} \sum_{\boldsymbol{m}}  E[Y_{\boldsymbol{i}\boldsymbol{r}}| A=1, \boldsymbol{c},\boldsymbol{l}, \boldsymbol{m}] P(\boldsymbol{i}|A=0, \boldsymbol{c})P(\boldsymbol{r}|A=1, \boldsymbol{c},   \boldsymbol{l}) P(\boldsymbol{l}|A=1, \boldsymbol{c})  P(\boldsymbol{c}|A=1) \\
& {\text{conditional independence}} \quad Y_{m_1, \ldots, m_K}\independent (M_1, \ldots, M_K)|\{A=1, \boldsymbol{C}, \boldsymbol{L}\}  \nonumber \\
&=\sum_{\boldsymbol{c}} \sum_{\boldsymbol{l}} \sum_{\boldsymbol{m}}  E[Y| A=1, \boldsymbol{c},\boldsymbol{l}, \boldsymbol{m}] P(\boldsymbol{i}|A=0, \boldsymbol{c})P(\boldsymbol{r}|A=1, \boldsymbol{c},   \boldsymbol{l}) P(\boldsymbol{l}|A=1, \boldsymbol{c})  P(\boldsymbol{c}|A=1) \\
& {\text{consistency \& positivity}}.  \nonumber 
\end{align}
}
}

\footnotetext[1]{Using short notation; for instance, $E[Y_{\boldsymbol{i}\boldsymbol{r}}| A=1, \boldsymbol{C}=\boldsymbol{c},\boldsymbol{L}=\boldsymbol{l}]=E[Y_{\boldsymbol{i}\boldsymbol{r}}| A=1, \boldsymbol{c},\boldsymbol{l}]$. For simplicity, the discrete case is considered; if continuous variables are involved, the transition would be to integration instead of summation.}

For $E[M_0]$ in $\text{IE}_{M}=E[Y_{1}-Y_{1 E[M_{0}]}]$ (equation \ref{IEM}), we have: $E[M_0]=E[M_0|A=0]$ by the independence assumption $M_a\independent A$, and $E[M_0|A=0]=E[M|A=0]$ by consistency and positivity. 

\section{MC g-computation for the estimation of \texorpdfstring{$E[Y_{a \boldsymbol{\widetilde{I}}_{a^*|\boldsymbol{C}} \boldsymbol{\widetilde{R}}_{a| \boldsymbol{C}}}]$}{E[Y]}}}

\begin{table}[H]
{
\renewcommand{\arraystretch}{1.1}
\begin{tabular}{p{0.095\textwidth}|p{0.875\textwidth}}
\hline
\multicolumn{2}{l}{\textbf{Algorithm to estimate $E[Y_{a \boldsymbol{\widetilde{I}}_{a^*|\boldsymbol{C}} \boldsymbol{\widetilde{R}}_{a| \boldsymbol{C}}}]$}}\\
\hline
     Step 1 & Specify and fit models for:\\
     & 1) the distribution of $\boldsymbol{L}$ conditional on $A$ and $\boldsymbol{C}$ (Model 1),\\
     & 2) the distribution of $\boldsymbol{R}$ conditional on  $A$ and $\boldsymbol{C}$ (Model 2),\\
        & 3) the distribution of $\boldsymbol{I}$ conditional on $A=a^*$ and $\boldsymbol{C}$ (Model 3),\\
         & 4) the outcome $Y$ conditional on $A$, $\boldsymbol{M}$, $\boldsymbol{L}$ and $\boldsymbol{C}$ (Model 4). \\
     Step 2 & Set the distribution of $\boldsymbol{C}$ to the respective empirical distribution. \\
      Step 3 & Set A=a. \\
    Step 4 & Set the variables in $\boldsymbol{L}$ to a random draw from Model 1 fitted in Step 1.\\
    Step 5 & Set the variables in $\boldsymbol{R}$ to a random draw from Model 2 fitted in Step 1.\\
    Step 6 & Set the variables in $\boldsymbol{I}$ to a random draw from Model 3 fitted in Step 1.\\
    Step 7 & Draw the outcome from Model 4 fitted in Step 1 based on the updated data generated through steps 2-6. \\
     Step 8 & Estimate $E[Y_{a \boldsymbol{\widetilde{I}}_{a^*|\boldsymbol{C}} \boldsymbol{\widetilde{R}}_{a| \boldsymbol{C}}}]$ by calculating the mean of $Y$. \\
      Step 9 & Repeat steps 4-8 Z times with different seeds and calculate the mean of the Z estimates (to reduce Monte Carlo error).\\
    Step 10 & To obtain the nonparametric bootstrap {standard error} and percentile confidence intervals, repeat steps 1-9 $B$ times on the bootstrapped data. \\
    \hline
\end{tabular}
}
\caption{Proposed algorithm to estimate $E[Y_{a \boldsymbol{\widetilde{I}}_{a^*|\boldsymbol{C}} \boldsymbol{\widetilde{R}}_{a| \boldsymbol{C}}}]$,  based on the estimator described by \citet{Daniel2017}.}
\label{Algo_IIE}
\end{table}

\section{Application to the gender pay gap}
\subsection{Structural assumptions}
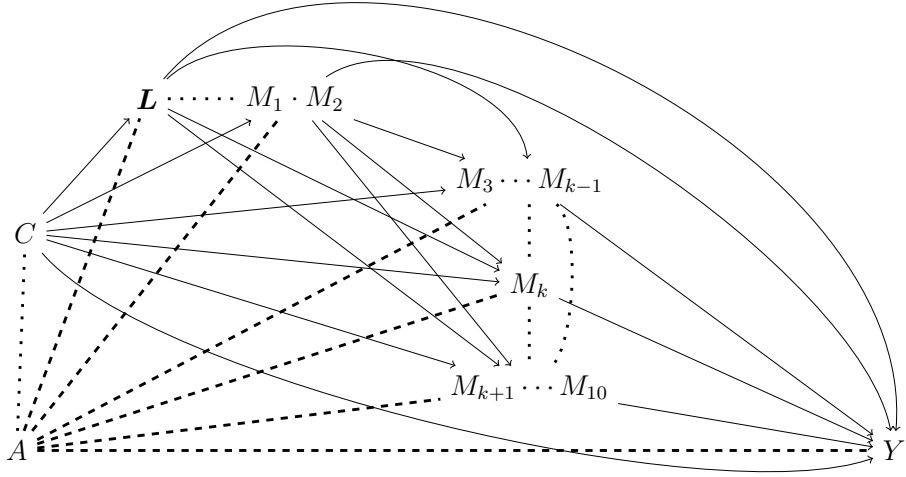
\begin{figure}[H]
\centering
 \scalebox{0.9}{
\begin{tikzpicture}[x=10in,y=6in]
\node (M1) at (0.16,-0.050) {$M_3 \cdots M_{k-1}$};
\node (A) at (-0.135,-0.310) {$A$};
\node (M7) at (0.16,-0.150) {$M_k$};
\node (M10) at (0.16,-0.250) {$M_{k+1} \cdots M_{10}$};
\node (Y) at (0.37,-0.310) {$Y$};
\node (C) at (-0.13,-0.1) {$C$};
\node (E1) at (0.025,0.03) {$M_1 \cdot M_2$};
\node (L) at (-0.06,0.03) {$\boldsymbol{L}$};

\draw [dashed, very  thick, -] (A) edge (M1);

\draw [dashed, very thick,-] (A) edge (M7);
\draw [dashed, very thick,-] (A) edge (Y);
\draw [dashed, very thick,-] (A) edge (M10);
\draw [->] (M1) edge (Y);
\draw [->] (M7) edge (Y);
\draw [->] (M10) edge (Y);
\draw [->] (C) edge (M1);
\draw [->] (C) edge (M7);
\draw [->] (C) edge (M10);
\draw [->] (C) to [out=-50,in=-160, looseness=0.5] (Y);

\draw [dashed, very thick,-]  (A) edge (L);
\draw [dashed, very thick,-]  (A) edge (E1);

\draw  [->] (L) to [out=45,in=95, looseness=0.7] (M1);
\draw  [->]  (L) edge (M7);
\draw  [->] (L) edge (M10);
\draw  [->] (L) to [out=50,in=85, looseness=0.9] (Y);
\draw [->] (E1) to [out=35,in=100, looseness=0.6] (Y);
\draw [->]  (E1) edge (M1);
\draw [->]  (E1) edge (M7);
\draw [->]  (E1) edge (M10);
\draw [->]  (C) edge (E1);
\draw [->] (C) edge (L);
\draw [dashed, very thick,-]  (A) edge (C);
\draw [loosely dotted, very thick]  (L) edge (E1);
\draw [loosely dotted, very thick]  (M1) edge (M7);
\draw [loosely dotted, very thick] (M7) edge (M10);
\draw [loosely dotted, very thick] (M1) to [out=-40,in=45, looseness=0.5] (M10);

\end{tikzpicture}
}
\caption{Gender pay gap: Assumptions regarding the causal influences among variables for a specific time period (here, the year 2017). Loosely dotted lines among the potential mediators signify that the structural dependence between these mediators remains unspecified. Dashed lines from $A$ (female/male category) to other factors emphasize that causal effects of $A$ are not addressed in the application of Approach 3.}
\label{DAG_applied}
\end{figure}
\subsection{Situating results within Approaches 1-3: Conditions and suggested interpretation}
\label{Contextualizing Aims}

The following Tables \ref{Contextualizing Aims 1a-b} and \ref{Contextualizing Aims 2a-b} summarize the essential conditions under which the effects obtained for Aims 1a and 1b, as well as for Aims 2a and 2b of the applied analysis described in the main text (Sections \ref{Applied analysis: Results Aim1} and \ref{Applied analysis: Results Aim2}), may represent specific target quantities within Approaches 1-3 (Sections \ref{Approach1}-\ref{Approach3}). At this point, it is up to the reader to decide whether or not causal effects related to gender are meaningful in this context. As a reminder, the mediators $M_1$ and $M_2$ represent education levels ($M_1$: $\geq$ 12 years of education, $M_2$: college degree or higher). They are assumed to precede the other mediators, which include job characteristics and work experience (see Figure \ref{DAG_applied}). In Aim 1a, $\boldsymbol{I} = (M_3, \ldots, M_{10})$, with $\boldsymbol{R} = (M_1, M_2)$. In Aim 1b, a separate analysis is conducted for each mediator $M_j$, where $\boldsymbol{I} = M_j$ for $j=3,\ldots,10$. In Aim 2a, the analysis considers all mediators together, with $\boldsymbol{I} = (M_1, \ldots, M_{10})$ and $\boldsymbol{R} = \emptyset$. In Aim 2b, a separate analysis is conducted for each mediator $M_k$, where $\boldsymbol{I} = M_k$ for $k=1,\ldots,10$. \(C\) denotes age in years, \(L_1\) indicates the presence of at least one child in the household, and \(L_2\) indicates having a direct migration background. $A$ is a binary indicator labeled ``Gender'', where $A=1$ denotes being female and $A=0$ denotes being male. It is assumed that $A$ is not affected by the other variables in the model. In Aims 1a and 1b, the intervention distribution of \(\boldsymbol{I}\), from which the random draws $\widetilde{\boldsymbol{I}}$ are obtained, is specified as 

\[
P_{\boldsymbol{I}|A=0, C, M_1, M_2}(\boldsymbol{i}|A=0, C, M_1, M_2).
\]
In both Aims 2a and 2b, the intervention distribution of $\boldsymbol{I}$ is specified as that observed in the total sample population: 

\[
P_{\boldsymbol{I}}(\boldsymbol{i}).
\]
For all Aims, the joint distribution of the other mediators, summarized in \(\boldsymbol{R}\), is held constant, given \(C\) and \(\boldsymbol{L}=(L_1, L_2)\). For notational simplicity, let GPG represent the observed marginal gender pay gap, defined as $E[Y|A=1] - E[Y|A=0]$. Let TE denote the total effect on wage of being assigned to the female category, defined as $E[Y_1 - Y_0]$.

\begin{table}[H] 
    \centering
{\textbf{Aims 1a-b}}\\[1ex] 
\centering
    \begin{adjustbox}{width=1\textwidth}
    \small
    \begin{tabularx}{\textwidth}{|l|p{1.3cm}|p{4.5cm}|p{3.8cm}|p{4.4cm}|}
        \hline
        \textbf{A} & \textbf{Inter\-vention on} & \textbf{Causal Estimand} & \textbf{Assumptions} & \textbf{Interpretation} \\
        \hline
        1 & 
    \small{Gender, Mediators} & 
         \makecell{$\frac{E[Y_1 - Y_{1\widetilde{\boldsymbol{I}}\widetilde{\boldsymbol{R}}_{1|C,\boldsymbol{L}_1}}]}{\text{TE}} \times 100$}
         
        \makecell{\\$\widetilde{ \boldsymbol{I}}$ from \\\small{$P(\boldsymbol{I}_0=\boldsymbol{i}| A=0, C, M_1, M_2)$}} 
        \makecell{\\$\widetilde{\boldsymbol{R}}$ from\\} 
        \small{$P(\boldsymbol{R}_1=\boldsymbol{r}| A=1, C, \boldsymbol{L})$
       } & 
        Consistency, Positivity,
        \makecell{\small{$Y_{a\boldsymbol{m}} \independent A$},\\ \small{$Y_{a\boldsymbol{m}} \independent \boldsymbol{M}|A=a, C, \boldsymbol{L}$},\\ \small{$\boldsymbol{M}_{a}  \independent A$} \citep{Daniel2017},\\}No unmeasured confounders of $\boldsymbol{I}$ and $M_1$, $M_2$ given \makecell{$A$ and $C$,\\}Indirect effect measure criteria \citep{Miles_2023}, \hspace{0.5cm} Correct model specification & 
       \% of TE mediated by all pathways from being assigned to the female category to wage through $\boldsymbol{I}$, except for those that pass through $M_1$ or $M_2$ (which precede $\boldsymbol{I}$), and those that pass through descendants of $\boldsymbol{I}$ in $\boldsymbol{R}$. \\
        \hline
2 & \small{Gender, Mediators} & \makecell{$\frac{E[Y_1 - Y_{1\widetilde{\boldsymbol{I}}\widetilde{\boldsymbol{R}}_{1|C,\boldsymbol{L}_1}}]}{\text{TE}} \times 100$}
         
        \makecell{\\$\widetilde{ \boldsymbol{I}}$ from\\ $P(\boldsymbol{I}_0=\boldsymbol{i}| A=0, C, M_1, M_2)$}
        \makecell{\\$\widetilde{\boldsymbol{R}}$ from\\} 
        \small{$P(\boldsymbol{R}_1=\boldsymbol{r}| A=1, C, \boldsymbol{L})$} & 
        Consistency, Positivity, \makecell{\small{$Y_{a\boldsymbol{m}} \independent A$},\\ \small{$Y_{a\boldsymbol{m}} \independent \boldsymbol{M}|A=a, C, \boldsymbol{L}$},\\\small{$\boldsymbol{M}_{a} \independent A$} \citep{Daniel2017},\\}Correct model specification  & \% of TE reduced by setting the counterfactual distribution of $\boldsymbol{I}$ under the female  category equal to that under the male category, given $C$ and $M_1$, $M_2$, while holding the counterfactual distribution of $\boldsymbol{R}$ under the female category constant, given $C$ and $\boldsymbol{L}_1$.\\
\hline
3 &\small{Media\-tors}& \makecell{$\frac{
E\left[Y- Y_{\widetilde{\boldsymbol{I}}\widetilde{\boldsymbol{R}}|A=1, C,\boldsymbol{L}} \middle| A=1\right]
}
{
\text{GPG}
} \times 100$} 
\makecell{\\
$\widetilde{\boldsymbol{I}}$ from \\
$P(\boldsymbol{I}=\boldsymbol{i}|A=0, C, M_1, M_2)$\\
\\$\widetilde{\boldsymbol{R}}$ from\\
$P(\boldsymbol{R}=\boldsymbol{r}|A=1, C, \boldsymbol{L})$} &Consistency, Positivity, \small{$Y_{a\boldsymbol{m}}\independent \boldsymbol{M}|A=a, C, \boldsymbol{L}$} \citep{Jackson}, Correct model specification & \% of GPG  reduced by setting the distribution of $\boldsymbol{I}$ among women equal to that observed among  men, given $C$ and $M_1, M_2$, while holding the distribution of $\boldsymbol{R}$ among women constant, given $C$ and $\boldsymbol{L}$.\\
        \hline
    \end{tabularx}
      \end{adjustbox}
    \caption{Targets of interventions, causal estimands, required assumptions, and interpretation when situating the results obtained for Aims 1a and 1b within Approaches ($\textbf{A}$) 1-3. GPG denotes the observed marginal gender pay gap, defined as $E[Y|A=1] - E[Y|A=0]$, TE denotes the total effect of being assigned to the female category on wage, defined as $E[Y_1-Y_0]$.}
    \label{Contextualizing Aims 1a-b}
   
\end{table}

\begin{table}[H] 
    \centering
{\textbf{Aims 2a-b}}\\[1ex] 
\centering
    \begin{adjustbox}{width=1\textwidth}
    \small
    \begin{tabularx}{\textwidth}{|l|p{1.3cm}|p{5.35cm}|p{3.65cm}|p{3.7cm}|}
        \hline
        \textbf{A} & \textbf{Inter\-vention on} & \textbf{Causal Estimand} & \textbf{Assumptions} & \textbf{Interpretation} \\
        \hline
2 & \small{Gender, Mediators} & \makecell{$1\;-\;$\\$\frac{E[Y_{1\widetilde{\boldsymbol{I}}\widetilde{\boldsymbol{R}}_{1|C, \boldsymbol{L}_1}}]\;-\;E[Y_{0\widetilde{\boldsymbol{I}}\widetilde{\boldsymbol{R}}_{0|C, \boldsymbol{L}_0}}]}{\text{TE}}$}
$(\times 100)$
\makecell{\\\\$\widetilde{\boldsymbol{I}}$ from $P_{\boldsymbol{I}}(\boldsymbol{i})$\\ (observed distribution of $\boldsymbol{I}$ \\in full sample)}
& 
        Consistency, Positivity, \makecell{$Y_{a\boldsymbol{m}}\independent A$, \\$Y_{a\boldsymbol{m}}\independent \boldsymbol{M}|A=a, C, \boldsymbol{L}$\\ \citep{robins1992identifiability, Pearl2001},\\
        $\boldsymbol{M}_{a}  \independent A$,\\}Correct model specification  & \% of TE reduced (``portion eliminated'' \citep{robins1992identifiability}) by setting the counterfactual distribution of $\boldsymbol{I}$ under the female and male  categories equal to the distribution in the full sample, while holding the counterfactual distribution of $\boldsymbol{R}$ constant, given $C$ and  $\boldsymbol{L}$.\\ 
\hline
3 &\small{Media\-tors          }& \makecell{$1\;-\;$\\$ \frac{E\left[Y_{\widetilde{\boldsymbol{I}} \widetilde{\boldsymbol{R}}|1, C, \boldsymbol{L}}| A=1\right]\;-\;
E\left[Y_{\widetilde{\boldsymbol{I}} \widetilde{\boldsymbol{R}}|0, C, \boldsymbol{L}} | A=0\right]}{\text{GPG}}$} 

$(\times 100)$
\makecell{\\$\widetilde{\boldsymbol{I}}$ from $P_{\boldsymbol{I}}(\boldsymbol{i})$\\ (observed distribution of $\boldsymbol{I}$ \\in full sample)} & Consistency, Positivity, \makecell{$Y_{a\boldsymbol{m}}\independent \boldsymbol{M}|A=a, C, \boldsymbol{L}$\\ \citep{Jackson},}Correct model specification & \% of GPG reduced by setting the distribution  of $\boldsymbol{I}$ among women and among men equal to the distribution in the full sample, while keeping the distribution of $\boldsymbol{R}$ among women and among men as observed, given $C$ and $\boldsymbol{L}$.\\
        \hline
    \end{tabularx}
      \end{adjustbox}
    \caption{Targets of interventions, causal estimands, required assumptions, and interpretation when situating the results obtained for Aims 2a and 2b within Approaches ($\textbf{A}$) 2-3. GPG denotes the observed marginal gender pay gap, defined as $E[Y|A=1] - E[Y|A=0]$, TE denotes the total effect of being assigned to the female category on wage, defined as $E[Y_1-Y_0]$.}
    \label{Contextualizing Aims 2a-b}
   
\end{table}

\subsection{Detailed results for Aims 1b and 2b}
\label{Aim1&2}
Aim 1b: \textit{Which intervention would yield the largest reduction when setting the distribution of a single mediator $M_j$ $(j=3,\ldots, 10)$ in women equal to that in men, given age and educational background, while keeping the joint distribution of the other mediators in women constant as observed, given $C$ and $\boldsymbol{L}$?}
 \begin{table}[H]
 \centering
 \label{Aim1b_Results}
\begin{tabular}{|l|r|}

\hline
Intervention target & \% Reduction in disparity in Y with  95\% CI  \\
\hline
 
  $M_3$: $\geq$ College degree job & 3.6 [1.3; 6.3]  \\
 $M_4$: $\geq 6.7$ years in company & 5.9 [4.6; 7.5] \\
 $M_5$: Female-dominated {industry}  & 7.4 [3.3; 11.3]\\
 $M_6$: Leading position & 7.4 [6.0; 9.6] \\
 $M_7$: Full-time employment & 6.3 [1.2; 11.4]\\
  $M_8$: Flexible working hours   & -2.3 [-3.9; -0.6]\\
 $M_9$: Work experience& 26.5 [20.9; 32.6] \\
 $M_{10}$: Job prestige (SIOPS)& 4.6 [2.3; 6.8]\\
 \hline
\end{tabular}
\caption{Aim 1b. Percentage reductions in gender pay gap in log gross hourly wages, achieved by single-mediator interventions in women, with 95\% bootstrap confidence intervals. A positive value indicates a reduction, while a negative value indicates an increase in the gender pay gap.}
\end{table}

Aim 2b: \textit{Which intervention would yield the largest reduction when setting the distribution of a single mediator $M_k$ $(k=1, \ldots, 10)$ in both women and men equal to that in the total sample population, while keeping the joint distribution of the other mediators in women and men constant as observed, given $C$ and $\boldsymbol{L}$?}
\begin{table}[H]
\centering
\resizebox{\columnwidth}{!}{

\begin{tabular}{|l|r|r|r|}

\hline
 & \% Reduction in &\% Reduction due to &\% Reduction due to\\
Intervention target & disparity in $Y$  & change in $Y$ in men & change in $Y$ in women \\
\hline
 $M_1$: $\geq 12$ years of education & -7.6\ [-14.9; 0.7] & -6.8\ [-13.8; 0.8] & -0.7\ [-3.1; 2.2]\\ 
$M_2$: $\geq$ College degree & -5.6\ [-14.1; 3.5] & -4.3\ [-12.3; 3.2] & -1.3\ [-5.3; 2.6]\\  
   $M_3$: $\geq$ College degree job & 8.6\ [0.3; 17.7] & 7.4 [2.4; 12.8] & 1.2\ [-5.0; 7.6] \\
 $M_4$: $\geq 6.7$ years in company & 4.2\ [-0.1; 8.2] & 3.8\ [-0.4; 6.7] & 0.3\ [-2.1; 2.8]\\
 $M_5$: Female-dominated {industry}  & 10.8\ [7.7; 14.1] & 7.8\ [5.8; 9.7] & 3.0\ [0.8; 5.6]\\
 $M_6$: Leading position &10.3\ [6.0; 14.2] & 4.5\ [1.3; 7.8] & 5.8\ [3.4; 8.2]\\
 $M_7$: Full-time employment & 29.8\ [20.2; 39.9]& 26.7\ [17.8; 36.4] & 3.1\ [0.5; 5.5]\\
  $M_8$: Flexible working hours   & 0.3\ [-2.7; 3.5] & 1.3\ [-1.2; 4.0]& -1.0\ [-2.7; 0.6]\\
 $M_9$: Work experience& 21.5\ [9.2; 35.3] & 4.7\ [-7.1; 16.0] & 16.8\ [9.6; 24.0]\\
 $M_{10}$: Job prestige (SIOPS)& 0.6\ [-4.8; 6.0]&-2.3\ [-5.1; 0.6] & 2.9\ [-1.8; 7.2]\\
 \hline
\end{tabular}%
}

\caption{Aim 2b. Percentage reductions in gender pay gap in log gross hourly wages, achieved by single-mediator interventions in men and in women, with 95\% bootstrap confidence intervals. A positive value indicates a reduction, while a negative value indicates an increase in the gender pay gap.}
\label{Aim2b_Results}
\end{table}

\end{document}